\documentclass[useAMS,usegraphicx,usenatbib]{mn2e}

\usepackage{color}

\def\degrees{^\circ}

\def\arcsec{^{\prime\prime}}

\def\kmsin{{\rm km s}^{-1}}

\def\kpc{{\rm kpc}}

\def\etal{{et al.\ }}
\def\eg{{\it e.g.\ }}

\def\cf{{\it cf.\ }}

\def\spose#1{\hbox to 0pt{#1\hss}}

\newcommand{\apprla}{\mathrel{\raise2pt\hbox{\rlap{\hbox{\lower6pt\hbox{$\sim$}}}\hbox{$<$}}}}
\newcommand{\apprga}{\mathrel{\raise2pt\hbox{\rlap{\hbox{\lower6pt\hbox{$\sim$}}}\hbox{$>$}}}}

\def\ud{{\mathrm d}}

\newcommand{\greeksym}[1]{{\usefont{U}{psy}{m}{n}#1}}

\newcommand{\uDelta}{\mbox{\greeksym{D}}}

\title[Dearth of dark matter or massive dark halo in NGC 3379?]{Dearth
  of dark matter or massive dark halo? Mass-shape-anisotropy
  degeneracies revealed by NMAGIC dynamical models of the elliptical
  galaxy NGC 3379}

\author[F. De Lorenzi \etal]{F. De
  Lorenzi$^{1,2}$\thanks{E-mail:lorenzi@mpe.mpg.de}, O. Gerhard$^2$,
  L. Coccato$^{2,3}$, M. Arnaboldi$^{4,5}$, M. Capaccioli$^6$, \and
  N.G Douglas$^3$, K.C. Freeman$^7$, K. Kuijken$^{8,3}$, M.R. Merrifield$^9$,
  N.R. Napolitano$^6$, \and E. Noordermeer$^9$,
  A.J. Romanowsky$^{10,3,9}$, V.P. Debattista$^{11}$ 
  \\ $^1$ Astron. Institut, Universit\"at Basel, Venusstrasse 7, Binningen,
  CH-4102, Switzerland 
  \\ $^2$ Max-Planck-Institut f\"ur Ex. Physik, Giessenbachstra{\ss}e, 
  D-85741 Garching, Germany 
  \\ $^3$ Kapteyn Astronomical Institute, University of Groningen, the
  Netherlands 
  \\ $^4$ ESO, Karl-Schwarzschild-Str. 2, D-85748 Garching Germany 
  \\ $^5$ INAF, Observatory of Turin, Strada Osservatorio 20,
  10025 Pino Torinese, Italy 
  \\ $^6$ Osservatorio di Capodimonte, Naples, Italy
  \\ $^7$ Research School of Astronomy \& Astrophysics, ANU, Canberra,
  Australia
  \\ $^8$ University of Leiden, The Netherlands
  \\ $^9$ School of Physics \& Astronomy, University of Nottingham, UK
  \\ $^{10}$ Departamento de F\'isica, Universidad de Concepsi\'on, Casilla 160-C,
  Concepci\'on, Chile
  \\ $^{11}$ Centre for Astrophysics, University of Central Lancashire, Preston,
  PR1 2HE, UK}

\begin{document}   
   
\date{Accepted ---. Received ---; in original form ---}

\pagerange{\pageref{firstpage}--\pageref{lastpage}} \pubyear{----}
  
\maketitle

\label{firstpage}

\begin{abstract} 

Recent results from the Planetary Nebula Spectrograph (PN.S) survey
have revealed a rapidly falling velocity dispersion profile in the
nearby elliptical galaxy NGC 3379, casting doubts on whether this
intermediate-luminosity galaxy has the kind of dark matter halo
expected in $\Lambda$CDM cosmology. We present a detailed dynamical
study of this galaxy, combining ground based long-slit spectroscopy,
integral-field data from the SAURON instrument, and PN.S data
reaching to more than seven effective radii.

We construct dynamical models with the flexible
$\chi^2$-made-to-measure particle method implemented in the NMAGIC
code. We fit spherical and axisymmetric models to the photometric
and combined kinematic data, in a sequence of gravitational
potentials whose circular velocity curves at large radii vary
between a near-Keplerian decline and the nearly flat shapes
generated by massive halos.

Assuming spherical symmetry we find that the data are consistent both
with near-isotropic systems dominated by the stellar mass, and with
models in massive halos with strongly radially anisotropic outer parts
($\beta\apprga 0.8$ at $7R_e$). Formal likelihood limits would exclude
(at $1\sigma$) the model with stars only, as well as halo models with
$v_{circ}(7R_e)\apprga 250\kmsin$. A sequence of more realistic
axisymmetric models of different inclinations and a small number
of triaxial tests confirm the spherical results.  All valid models
fitting all the data are dynamically stable over Gyrs, including the
most anisotropic ones.

Overall the kinematic data for NGC 3379 out to $7R_e$ are consistent
with a range of mass distributions in this galaxy. NGC 3379 may well
have a dark matter halo as predicted by recent merger models within
$\Lambda$CDM cosmology, provided its outer envelope is strongly
radially anisotropic.

\end{abstract}


\section{Introduction}   
\label{sec:introduction}
There is strong evidence that most galaxies are surrounded by massive
dark matter (DM) halos. This is most evident in spiral galaxies, where
the rotation curves of extended cold gas disks remain flat out to
large radii.  In elliptical galaxies the evidence for dark halos has
built up more slowly, and their halo properties are not so well known,
because of a lack of ubiquitous tracer similar to the HI rotation
curves in spirals. Only in a few cases is it possible to measure masses
from extended HI ring velocities \citep[\eg][]{franx+94,oosterloo+02}.

However, at least for giant elliptical galaxies stellar-dynamical
studies from integrated light spectra
\citep[\eg][]{krona_etal00,gerhard+01, cappellari+06, thomas+07},
analyses of the X-ray emitting hot gas atmospheres
\citep[\eg][]{awaki+94,matsushita+98,loewenstein+99,humphrey+06,fukazawa+06},
and gravitational lensing data \citep[\eg][]{wilson+01, treu+koop04,
  rusin+kochanek05,koopmans+06,gavazzi+07} are now giving a fairly
consistent picture.  The general result from these studies is that
these ellipticals are surrounded by dark matter halos, the inferred
mass profiles (luminous plus dark) are nearly isothermal, i.e., the
circular velocity curves approximately flat, and the dark matter
contributes $\sim10-50\%$ of the mass within $R_e$. The central DM
densities in ellipticals are higher than in spirals, presumably
reflecting their earlier formation epochs \citep{gerhard+01,thomas06}.

In light of this, the finding of \citet{romanowsky_etal03} and 
\citet{douglas+07},
that several intermediate luminosity ellipticals (NGC 3379, NGC 4494,
NGC 821) apparently have only diffuse dark matter halos if any, is
quite surprising.  Could the dark matter properties of these
ellipticals be different from those of giant ellipticals
\citep[e.g.,][]{napolitano+05}, perhaps related to the fact that these
lower-luminosity galaxies are less often found in groups or clusters?
The result of \citet{romanowsky_etal03} is based on the outer velocity
dispersion profiles of the three galaxies, determined from individual
planetary nebulae (PNe) velocities measured with the special PN.S
instrument \citep{douglas+02}.  Two of the three galaxies are nearly
round on the sky, and therefore the dynamical analysis was carried out
with spherical models.  A fourth galaxy with a fairly rapidly
declining outer velocity dispersion profile is NGC 4697
\citep{mendez_etal01}. However, using axisymmetric particle models
\citet{delo+08} have recently shown that only models with massive 
dark halos are consistent with all the kinematic data for this galaxy,
and that the best models have circular velocity $v_c(5R_e)\simeq
250\kmsin$ at 5 effective radii.  Unfortunately, the diffuse gas
envelopes of these intermediate luminosity ellipticals have very low
densities, so an independent confirmation with X-ray data is
difficult.

The results of \citet{romanowsky_etal03} were criticized by
\citet{dekel+05}. These authors pointed out that the well known
mass-anisotropy degeneracy in the study of velocity dispersion
profiles does not allow one to unambiguously determine the mass
profile, that the triaxial nature of elliptical galaxies can cause low
line-of-sight velocity dispersions at some viewing angles, or that the
PNe could trace young stars generated during the merger formation
instead of the bulk of the old stars as usually assumed.
\citet{douglas+07} argued that \citet{romanowsky_etal03} properly took
into account orbital anisotropies in the data fitting process, that
the effect of triaxiality is very unlikely to be present in all three
galaxies, that the PN number density and velocity dispersion profiles
match the corresponding integrated light profiles reasonably well, and
that this as well as the universality of the bright end of the PN
luminosity function rules out that PNe only trace a young stellar
population. \citet{douglas+07} concluded that their results continue
to conflict with the presence of dark matter halos as predicted in
cosmological merger simulations.

The issue is important enough to merit a further careful analysis. In
this paper we construct dynamical models of NGC 3379 with the very
flexible NMAGIC particle code, making use of a variety of kinematic
data, including SAURON integral-field data, slit kinematics, and the
PN dispersion profile. 

The NMAGIC method is flexible not only with regard to anisotropy, but
also in allowing axisymmetric or triaxial shapes with radially varying
axis ratios. This is important since the intrinsic shape of NGC 3379
is still in doubt.  \citet{capaccioli+91} and \citet{statler_smha99}
have suggested that NGC 3379 is a triaxial S0 galaxy seen almost
face-on.  \citet{statler01} considered triaxial dynamical models and
constrained the shape of this galaxy to be axisymmetric and oblate in
the inner parts and triaxial in the outer parts.
\citet{shapiro_etal06} argue that the most likely model is one of a
moderately inclined oblate system.

The outline of the paper is as follows. In Section \ref{sec:obsdata}
we describe briefly how the various observational data for NGC 3379
are used in the modelling. In Section~\ref{sec:nmagic} we give a few
details of the $\chi^2$M2M NMAGIC method, and show how it performs on
a mock galaxy data set similar to that for NGC 3379. In
Section~\ref{sec:models} we then construct various dynamical models
for the real galaxy data, both spherical and flattened, in a sequence
of potentials with increasing circular velocity at large radii. As
summarized in the final Section~\ref{sec:conclusions} of the paper,
our main conclusion is that the combined kinematic data for NGC3379
is consistent with a range of dark matter halos including those
found by \citet{dekel+05} in their cosmology-based merger
simulations.

\section{Observational Data}
\label{sec:obsdata}
We begin by describing the observational data used in this study,
which are all taken from the literature. We also describe here the
procedure employed for obtaining the three-dimensional luminosity
density from the surface brightness data. In the following we adopt a
distance $9.8 \;\mathrm{Mpc}$ to NGC 3379 \citep{jensen_etal03},
effective radius $R_{e}=47\arcsec$ ($2.23\;\mathrm{kpc}$), and an
absolute B magnitude $M_B=-19.8$ \citep{douglas+07}.
\subsection{Photometric Data}
\label{sec:photometricdata}
\begin{figure}
\centering 
\includegraphics[width=0.95\hsize,angle=-90.0]{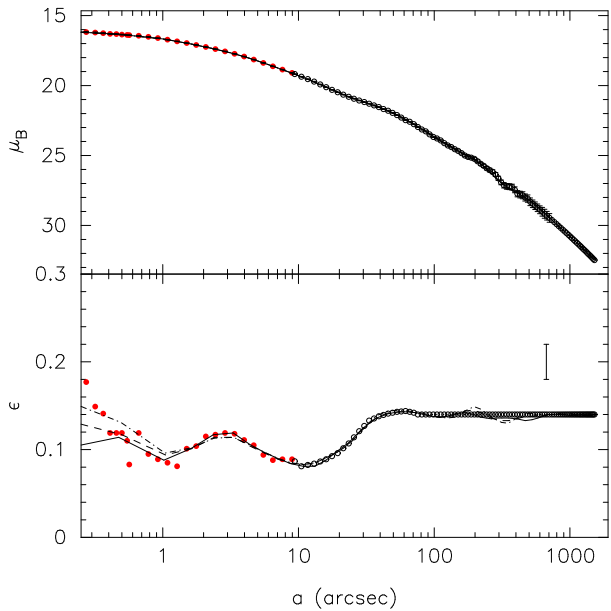}
\vskip0.3truecm
\caption[]{Combined photometry of NGC 3379 from \citet{capaccioli+90}
  (open black circles) and \citet{gebhardt+00} (full red circles). The
  two panels show the surface brightness (SB) profile and the
  ellipticity $\epsilon$ as a function of major axis distance.  Beyond
  $500\arcsec$ the SB points are from a Sersic model fitted to the
  interior data, and outside $81 \arcsec$, the ellipticity has been
  set to $\epsilon=0.14$.  In the ellipticity panel the error bar with
  size 0.02 illustrates the typical errors in the outer ellipticity
  measurements. The isophotal shape parameters $a_4$ and $a_6$ are not
  measured; they are set to zero.  The lines show three-dimensional
  luminosity models determined from these data and reprojected onto
  the sky, for assumed inclinations of $i=90\degrees$ (edge-on, full
  lines), $i=50\degrees$ (dashed lines), and $i=40\degrees$
  (dash-dotted lines).}
\label{fig:photo}
\end{figure}

The photometric data used in the present work consists of the
wide-field B-band photometry of \citet{capaccioli+90}, combined with
the HST V-band observations of \citet{gebhardt+00} to increase the
spatial resolution within the inner $10\arcsec$. The photometry has
been matched up by assuming a constant color offset $B-V=1.03$.
The last eight surface brightness (SB) points from
\citet{capaccioli+90}, outside $R\simeq500\arcsec$, show fluctuations
of an amplitude which we judged unphysical; these points we have
replaced with a \citet{sersic68} profile fitted to the galaxy further
in. The same Sersic fit is used to extrapolate the SB profile outside
the last measured point at $R=676\arcsec$. Similarly, we have replaced
the measured ellipticities for $R > 81\arcsec$, where the
observational uncertainties become large, by $\epsilon=0.14$.  Figure
\ref{fig:photo} presents the combined photometric data, showing
surface brightness and ellipticity $\epsilon$. The isophotal
shape parameters $a_4$ and $a_6$ are not available for these data and
are thus set to zero. For the spherical models, we have used the SB
profile rescaled to a mean radius $R_m \equiv
\sqrt{ab}=a\sqrt{1-\epsilon}$. For the axisymmetric models, we have
used a constant PA of $70\degrees$; the isophotal PA measured by
\citet{capaccioli+90} are within $\pm3\degrees$ of this value.
%

\subsection{Deprojection}
\label{sec:depro}
In our implementation of NMAGIC a particle model can be fitted to the
surface brightness and/or the deprojected luminosity density, \cf
Section \ref{sec:nmagic}. Below we use both options, so first need to
construct models for the three-dimensional luminosity density, $j$.

In the spherical case the surface brightness can be deprojected
uniquely.  For an axisymmetric system the deprojection is unique only
for edge-on galaxies; for systems inclined at an angle $i$ with
respect to the line-of-sight, the SB map contains information about
the luminosity density only outside a ``cone of ignorance'' in Fourier
space, of opening angle $90\degrees-i$, when $i=90\degrees$ denotes
edge-on \citep{rybicki87}. Thus, the deprojection of a moderately
inclined galaxy results in undetermined konus densities
\citep{gerhard_bin96,romanowsky_kochanek97}.

We deproject the surface brightness of NGC 3379, without correcting
for PSF effects, using the program of \cite{magorrian99}. The program
finds a smooth axisymmetric density distribution consistent with the
SB distribution for the specified inclination angle, by imposing that
the solution maximizes a penalized likelihood. This ensures that the
shape of the 3D luminosity density is smooth and biases the model
towards a specified diskyness. Because of the disklike nature of the
undetermined konus densities, requesting the luminosity density to
have zero diskyness effectively chooses between the different density
distributions that fit the SB data for $i\ne 90\degrees$. We have
used the program to compute luminosity densities for NGC 3379 for the
inclinations $i=90\degrees$, $i=50\degrees$ and $i=40\degrees$. Figure
\ref{fig:photo} compares the observed photometry with the three
deprojections reprojected onto the sky.  Varying the inclination from
$90\degrees$ to $40\degrees$ changes the intrinsic shape of the galaxy
from E1 to E3.

\subsection{Kinematic Data}
\label{sec:kindata}
\subsubsection{Stellar-absorption line data}
\label{sec:absndata}
\begin{figure}
\centering
\includegraphics[width=0.95\hsize,angle=-90.0]{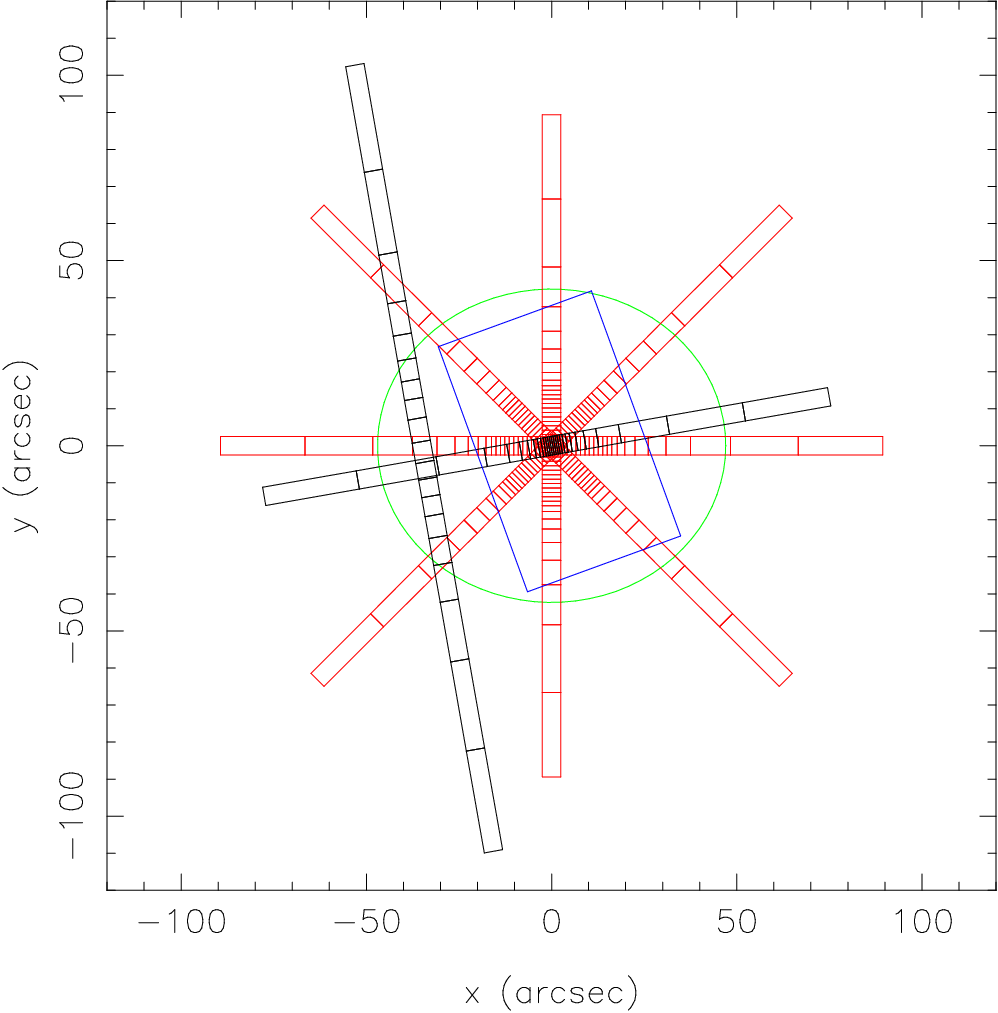}
\vskip0.5truecm
\caption[]{Schematic view of the positions with kinematic data as used
  to construct the dynamical models. The slits from
  \citet{statler_smha99} and \citet{krona_etal00} are coded in red and
  black, respectively. Boxes along the slits show the region of the
  galaxy for which respective kinematic data points were derived;
  these boxes are used to determine the luminosity-weighted
  Gauss-Hermite moments.  The blue rectangle indicates the SAURON
  field-of-view. The ellipse shown is oriented along PA=$70\degrees$,
  the average major axis of the photometry, and has a semi-major axis
  of length $R_{e}$ and axis ratio $q=0.9$.}
\label{fig:kinsetup}
\end{figure}

We have taken long-slit absorption line kinematics from the
literature. We use data from \cite{statler_smha99} at four different
position angles, extending out to radii of $\simeq 80\arcsec$.  We
complement these kinematics with the spectroscopic data from
\cite{krona_etal00}.  The major axis slits from \cite{statler_smha99}
and \cite{krona_etal00} are misaligned by $10\degrees$ in PA; however,
the data along both major axis slits follow each other closely. The
measurements along the shifted slit of \citet{krona_etal00} reach
$100\arcsec$ from the center.  From both kinematic data sets we have
the line-of-sight velocity, velocity dispersion, and higher order
Gauss-Hermite moments $h_3$ and $h_4$
\citep{gerhard93,vdmarel_franx93}.  Figure \ref{fig:kinsetup} shows
the schematic arrangement of the kinematic slits used in the dynamical
modeling.
%

\begin{figure}
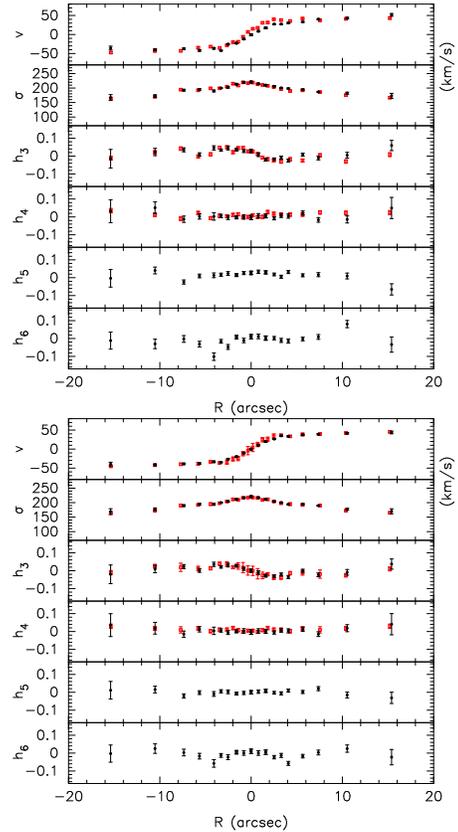

\centering
\includegraphics[width=0.65\hsize,angle=-90.0]{fig4a_krona_vs_sau.ps}
\includegraphics[width=0.65\hsize,angle=-90.0]{fig4b_symm_krona_vs_sau.ps}
\vskip0.5truecm
\caption[]{Comparison of the line-of-sight velociy distribution data
  along the galaxy's major axis ($PA=70^\circ$). The black circles
  correspond to the SAURON data and the open square symbols in red show
  the \cite{krona_etal00} data. The upper panel compares the original
  datasets, the lower panel is for the symmetrized data. In each panel
  from top to bottom are shown: $v$, $\sigma$, $h_3$, $h_4$, $h_5$,
  and $h_6$, for the latter two there are only SAURON data.}
\label{fig:krona_vs_sau}
\end{figure}

In addition to the long-slit kinematics we also use the integral-field
spectroscopy obtained with the SAURON instrument. These kinematic data
were kindly provided by \citet{shapiro_etal06} and consist of
line-of-sight velocity, velocity dispersion and higher order
Gauss-Hermite moments up to $h_6$. The SAURON field-of-view (FoV),
shown by the (blue) rectangle in Figure \ref{fig:kinsetup}, extends
from $-19.6\arcsec$ to $24.4\arcsec$ along its short boundary and from
$-34.8\arcsec$ to $35.6\arcsec$ along the long boundary.  In this FOV,
the positions of the 55x88 ``lenslets'' with which spectra were taken
define a fine grid of $4840$ grid cells, which serve as the basis grid
to define the $1602$ voronoi cells on which the final kinematic
measurements are given. This results in a total of $9612$ kinematic
SAURON observables, as well as $1602$ bin-luminosity observables.  The
SAURON data are  reproduced and compared to dynamical models in
Section \ref{sec:models}.  Each of the six panels shows the $1602$
voronoi bins, giving (from left to right) $v$, $\sigma$, $h_3$, $h_4$,
$h_5$ and $h_6$.
A comparison of the SAURON data with the data of \citet{krona_etal00}
along their major axis is given in Figure \ref{fig:krona_vs_sau}. Overall,
the two data sets agree well with each other. The same is true for
the comparison of the SAURON data with \citet{statler_smha99}, as shown
by \citet{shapiro_etal06}.

Both the SAURON data and the slit data are slightly asymmetric with
respect to the center of the galaxy. If we denote the original SAURON
dataset with $I(x,y|v_{\rm los},\sigma_{\rm los},h_3,h_4,h_5,h_6)$,
and with $I^\ast(x,y|v_{\rm los},\sigma_{\rm los},h_3,h_4,h_5,h_6) =
I(-x,-y,|-v_{\rm los},\sigma_{\rm los},-h_3,h_4,-h_5,h_6)$ the dataset
obtained from $I$ by point-symmetrical reflection with respect to the
origin, we can construct a symmetrized dataset $\bar{I}\equiv 0.5(I +
I^{\ast})$.  This symmetrized dataset $\bar{I}$ has a $\chi^2$ per
data point with respect to $I$ of $\chi^2/N=1.01$ when the original
errors are used.  Any point-symmetric model fit (spherical,
axisymmetric, triaxial) to the original data $I$ will therefore have a
systematic error floor of this magnitude. In the models below, we will
actually fit the symmetrized SAURON data to avoid any systematic
effects, but keep the original errors on both sides of the galaxy 
separately \citep[see also][]{shapiro_etal06}.

In a similar fashion, we have constructed symmetrized slit data sets.
To do this we average the two points at nearly similar radius on both
sides of the slit with respect to the center. Taking into account the
sign reversals of $v$ and $h_3$, we take for the symmetrized
data point the weighted mean of the points on both sides, with weights
proportional to the inverse square of the measurement errors, and
assign a new weighted error for the averaged point. If $\sigma_+$ and
$\sigma_-$ are the errors on both sides, the weights are
$w_+=1/\sigma_+^2$, $w_-=1/\sigma_-^2$, and the new error $\sigma$ is
given by the maximum of $2/\sigma^2=1/\sigma_+^2+1/\sigma_-^2$ and
half of the deviation between the original data points on both sides.
Again, the symmetrized data have a $\chi^2/N=1.0$ systematic deviation
from the original data, and therefore we will fit the symmetrized data
below to avoid the model being pulled around by points with small
error bars but large systematic deviations.
The second panel of Figure \ref{fig:krona_vs_sau} compares the
symmetrized SAURON data with the symmetrized \cite{krona_etal00} data
along the same slit as before. Again, the two data sets agree well
with each other.
%
\subsubsection{PNe data}
\label{sec:pnedata}
Planetary nebulae (PNe) are dying low- to intermediate mass stars that
emit most of their light in a few narrow lines of which the
$[\mathrm{OIII}]\lambda 5007$ is the most prominent one. Because there
are hardly other emission sources in elliptical galaxies, they can be
detected fairly easily, and once identified, their line-of-sight
velocity can be estimated from the Doppler shift of the emission line.
The PN population in elliptical galaxies is expected to arise from the
underlying galactic population of old stars and hence the PNe can be
used as kinematic tracers for the stellar distribution. Their number
relative to the luminosity of the galaxy is parametrized by the
$\alpha$ parameter, which is a function of colour \citep{hui_etal95}.

\citet{douglas+07} processed observations of NGC 3379 conducted with
the Planetary Nebula Spectrograph (PN.S) instrument and detected 214
spatially and spectrally unresolved PN candidates of which 191 are
assigned to NGC 3379.  Using the ``friendless'' algorithm applied by
\citet{merrett+03} they identified a small number of velocity
outliers, probably unresolved background galaxy contaminants, which
would be uniformly spread in velocity. The algorithm determined that
$2$ emission objects were more than $n=5$ standard deviations away
from the centroid of the velocity distribution of their $N=15$ nearest
neighbours, and  $3$ objects more than $3$ standard deviations
$\sigma$ \citep[see Fig.~8 of][]{douglas+07}.  The $3\sigma$ line
itself has considerable uncertainty at large radii, due to the small
number of PNe found there. Thus the exclusion of the outermost outlier
is somewhat uncertain. Because this object does have some influence on
the outermost velocity dispersion point, we will compare the models to
the data obtained both with and without this PN.

The radial distribution of the PNe in NGC 3379 was found to be
consistent with the stellar density profile, and their kinematics
consistent with absorption-line data in the region where the data sets
overlap.  Because the kinematics of the PNe in NGC 3379 are dominated
by random motions with little azimuthal variation, the velocity
dispersion can be computed in radial annuli without losing significant
dynamical structure. We will thus use the radial run of the
azimuthally averaged PN velocity dispersion in the dynamical
modelling, but also compare the models to the individual velocities in
a relative likelihood sense (cf.\ the tables and figures in Section
\ref{sec:models}).
\section{NMAGIC modelling}
\label{sec:nmagic}
To investigate the amount of dark matter consistent with the
kinematic data for NGC 3379, we construct a range of dynamical models
for the stellar component of this galaxy. We use the flexible
$\chi^2$-made-to-measure ($\chi^2$M2M) particle method as described
and implemented in the NMAGIC code by \citet{delo+07, delo+08}.
$\chi^2$M2M is a development of the M2M algorithm of \citet{ST96} that
is suitable for modelling observational data. The M2M methods work by
gradually adjusting individual particle weights as the model evolves,
until the N-particle system reproduces a set of target constraints.
In $\chi^2$M2M the standard $\chi^2$ statistics is used in the
function to be maximized upon convergence of the weights. This allows
for a proper treatment of observational errors, and the quality of the
final model can be assessed directly from the target data.

Compared to the familiar Schwarzschild method
\citep[\eg][]{schwarzschild79,rix_etal97,vdmarel+98,cretton_etal99,
romano_koch01,gebhardt_etal03,thomas+04, valluri_etal04,cappellari+06,
chaname+07, van_den_bosch+08} the particle approach is relatively new
and there are as yet only a few galactic dynamics studies in which it
has been employed.
\citet{bissantz_etal04} made a first practical application of the M2M
method of \citet{ST96} and constructed a dynamical model of the Milky
Way's barred bulge and disk by constraining the projected density
map. First attempts to extend the M2M method to account for kinematic
observables in addition to density constraints were made by
\citet{delo_etal06} and \citet{jourdeuil+ems06}.  However, a proper
treatment of observational errors was not yet included in their
implementations. \citet{delo+07} incorporated this in their
$\chi^2$M2M algorithm and demonstrated the potential of the NMAGIC
code by constructing particle models for spherical, axisymmetric,
triaxial and rotating target stellar systems. Some extensions of the
method and the first detailed modelling of slit kinematic and PN data
for an elliptical galaxy (NGC 4697) are described in \citet{delo+08}.

The NMAGIC method is flexible not only with regard to the orbit
structure, but also in allowing axisymmetric or triaxial shapes with
varying axis ratios.  Contrary to Schwarzschild's method, the
``best'' stellar density and luminous potential need not be specified
beforehand, but can be found from the evolution of the model. This
makes it ideal for the present study because different intrinsic
shapes have been suggested for NGC 3379 (see the Introduction) and the
issue of whether the kinematics require or allow dark matter may well
be connected not only with the orbital anisotropies but also with the
detailed shape of the stellar density distribution of the galaxy.
Given that NGC 3379 is nearly round on the sky, we have constrained
most models in this paper to be axisymmetric, with density
distribution fixed from the deprojection; however, for some models (in
Sections \ref{sec:oblate_spot} and \ref{sec:noaxi}) the stellar system
evolves towards a final, "best" density distribution, allowing for
radial variations in axis ratio. This approach has proved sufficient
for answering our main science question.

\subsection{Luminous and dark mass distributions}
\label{sec:mass}
As in \citet{delo+08}, we assume that the luminous mass of NGC 3379
follows the light and characterize it by a constant
mass-to-light ratio $\Upsilon$, so that the stellar mass density is
given by $\rho_{\star}=\Upsilon j$.  The total gravitational
potential is generated by the combined luminous mass and dark matter
distributions, $\phi = \phi_{\star}+\phi_L$, where $\phi_{\star}$ is
generated by $\rho_{\star}=\Upsilon j$. Only the luminosity density
$j$ is represented by the $N$-particle system. Its potential is
computed using a spherical harmonic decomposition as described in
\cite{sellwood03,delo+07}. The stellar potential is allowed to vary
during the modeling process, but the DM halo is rigid.

Here our aim is not to determine the detailed shape of the dark matter
halo in NGC 3379, but rather to first see whether the PN velocities
allow or require any dark matter at all in this galaxy.  To answer
this question we will investigate a one-dimensional sequence of
potentials whose circular velocity curves vary at large radii between
the near-Keplerian decline expected when the mass in stars dominates,
and the nearly flat shapes generated by massive dark halos. As in
\citet{delo+08} we thus represent the dark matter halo by the
logarithmic potential \citep{bin_tre87}
\begin{equation}
\phi_L(r) = \frac{v_0^2}{2}\ln(r_0^2+r^2).
\label{eqn:logpot}
\end{equation}

\subsection{Model and target observables}
\label{sec:target}
Target observables include surface or volume densities and
line-of-sight kinematics. For modelling the luminosity distribution of
NGC 3379, we generally use the deprojected luminosity density of NGC
3379, expanded in spherical harmonic coefficients $A_{lm}$ on a 1-D
radial mesh of radii $r_k$. The corresponding model observables are
computed from the particles based on a cloud-in-cell (CIC) scheme; see
\citet{delo+07}.

In some models, we do not constrain the three-dimensional luminosity
density but only the stellar surface density, leaving the former free
to evolve. In the remaining cases, we constrain the model by both the
deprojected luminosity density and the projected surface density.  In
a similar spirit as for the volume density, we use as target
constraints for the observed SB distribution the coefficients of a
Fourier expansion in the azimuthal angle, computed on a 1-D radial
mesh of projected radii $R_k$. For the corresponding model
observables, the particles are assigned to the radial grid using a CIC
scheme, and the Fourier coefficients $a_m$ and $b_m$ for the particle
model on shell $k$ are computed via
\begin{equation}
a_{m,k}  =  L \sum_i \gamma_{ki}^{CIC} \cos( m\varphi_i) w_i
\label{eqn:fourier1}
\end{equation}
\begin{equation}
b_{m,k}  =  L \sum_i \gamma_{ki}^{CIC} \sin( m\varphi_i) w_i, \;\;m>0
\label{eqn:fourier2}
\end{equation}
where $w_i$ are the particle weights, $\varphi_i$ their angular
positions, and $\gamma_{ki}^{CIC}$ is a radial selection function.  We
use units for which the light $L_i$ of a stellar particle can be
written as $L_i = L w_i$ with $L$ the total luminosity of the galaxy.

As kinematic constraints, we use the luminosity-weighted Gauss-Hermite
coefficients from the SAURON or slit data, and luminosity-weighted
velocity moments for the PN data. For the SAURON data
\citep{shapiro_etal06}, the luminosity-weighted coefficients are
determined from the truncated Gauss-Hermite representation of the
line-of-sight velocity distribution (LOSVD) up to order $h_6$ and the
luminosity in the corresponding Voronoi bin. For the slit data
\citep{statler_smha99, krona_etal00}, they are constructed again from
the measured Gauss-Hermite moments, up to order $h_4$, and the
luminosity in the slit section corresponding to the relevant LOSVD.
The PN data \citep{douglas+07} are modelled either as 1-D radial
dispersion profile or as a discrete set of velocities; in the former
case we use as suitable observables the second velocity moments
$v^2_{los}$, luminosity-weighted by the number of PNe per radial bin.

The corresponding model observables $y_j$ are construced from the
particles via equations of the form
\begin{equation}
y_j(t)=\sum_{i=1}^N w_i K_j \left[ {\mathbf z}_i(t) \right],
\label{eqn:obs}
\end{equation}
where $w_i$ are the particle weights and ${\mathbf z}_i$ 
are the phase-space coordinates of the particles, $i=1,\cdots,N$.
Here the Kernel $K_j$ corresponds to the observable $y_j$. Detailed
expressions for the kinematic model observables are given in
\citet{delo+07, delo+08}.

In general, we replace the observables by the corresponding temporally
smoothed quantities to increase the effective number of particles in
the system, \cf \cite{ST96,delo+07}. For the parameters chosen, the
smoothing is typically over $\sim 10^3$ correction time steps.

\subsection{Constructing a particle model for the target data}
\label{sec:fits}
Generating an NMAGIC model for a set of observational constraints
proceeds by evolving the force-of-change (FOC) equations for the
particle weights,
\begin{equation}
\frac{\ud w_i(t)}{\ud t} = \varepsilon w_i(t)\left(\mu \frac{\partial
S}{\partial w_i} - \sum_j \frac{K_j \left[{\mathbf
z}_i(t)\right]}{\sigma(Y_j)} \uDelta_j(t) \right) 
\label{eqn:myFOC}
\end{equation} 
depending on the discrepancies between model ($y_j$) and target
observables ($Y_j$), $\uDelta_j(t) = (y_j-Y_j) / \sigma(Y_j)$. Here
$\sigma(Y_j)$ in the denominator is the error in the target
observable. Evolving the particle weights to convergence in this
way is equivalent to maximizing the merit function
\begin{equation}
F = \mu S-\frac{1}{2}\chi^2 
\label{eqn:F}
\end{equation}
with respect to the particle weights $w_i$, where for the profit
function $S$ we use the entropy, and the standard $\chi^2$ measures
the goodness of the fit. The parameter $\mu$ controls the contribution
of the entropy function to $F$.  The entropy term pushes the particle
weights to remain close to their priors, so models with large $\mu$
will have smoother distribution functions than those with small $\mu$.
The best choice for $\mu$ depends on the observational data to be
modeled, \eg spatial coverage and phase-space structure of the galaxy
under consideration, but also on the initial conditions, and will be
determined for the NGC 3379 dataset in the following Section
\ref{sec:isorot}.

Any NMAGIC model starts from a suitable initial model. For the
models presented in this paper, we have used as initial conditions a
\citet{hernquist90} model particle realization generated from a
distribution function (DF), using the method described in
\citet{victor_sell00}. The particle realization consists of $7.5
\times 10^5$ particles, has a scale length $a=1$, maximum radius
$r_{\rm max}=60$, and a total luminosity of unity. In model units, the
gravitational constant is $G=1$.  In real units, the model
  lengthscale corresponds to $50\arcsec$. Thus when we match the model
  to NGC 3379, the effective radius of NGC 3379 becomes 0.94 model
  units, or $2.23 \mathrm{kpc}$ at a distance of $9.8\mathrm{Mpc}$.

\subsection{Anisotropic mock galaxy model}
\label{sec:isorot}
To prepare for the modeling of NGC 3379, we now construct a spherical
mock galaxy model with known intrinsic properties to determine the
optimal value of the entropy ``smoothing'' parameter $\mu$ in
equation~(\ref{eqn:F}). Following a similar approach as in
\citet{gerhard_etal98} and \citet{thomas_etal05} we determine for
which value of $\mu$ the fitted particle model best reproduces the
intrinsic velocity moments of the input mock galaxy model. The
``best'' value of $\mu$ depends on the observational data to be
modelled and their spatial coverage, on the phase-space structure of
the galaxy, but also on the initial conditions from which the NMAGIC
modelling starts. The same value can then be used for the modelling of
NGC 3379, provided the mock galaxy is a reasonable approximation to
the real galaxy.

\begin{figure}
\centering
\includegraphics[angle=-90.0,width=0.95\hsize]{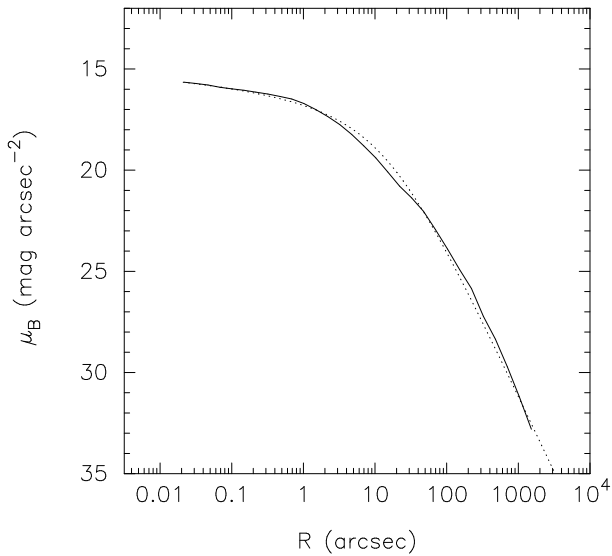}
\vskip0.2truecm
\caption[]{A comparison of the surface brightness profile of the
  mock galaxy model (dotted line) with that of NGC 3379 (full
  line), along the major axis of NGC 3379. }
\label{fig:gamma_vs_n3379}
\end{figure}

For the luminosity density of the mock galaxy we use a
\citet{hernquist90} model with total luminosity $L=1.24 \times
10^{10}\;L_{\odot,B}$ and scale radius $a=0.8 \;\kpc$, corresponding
to $R_{e} \approx 30''$ for the distance of NGC 3379.  As Figure
\ref{fig:gamma_vs_n3379} shows, the surface brightness profile of this
model galaxy  is a good approximation for NGC 3379.
%

\begin{figure}
\centering
\includegraphics[angle=-90.0,width=0.95\hsize]{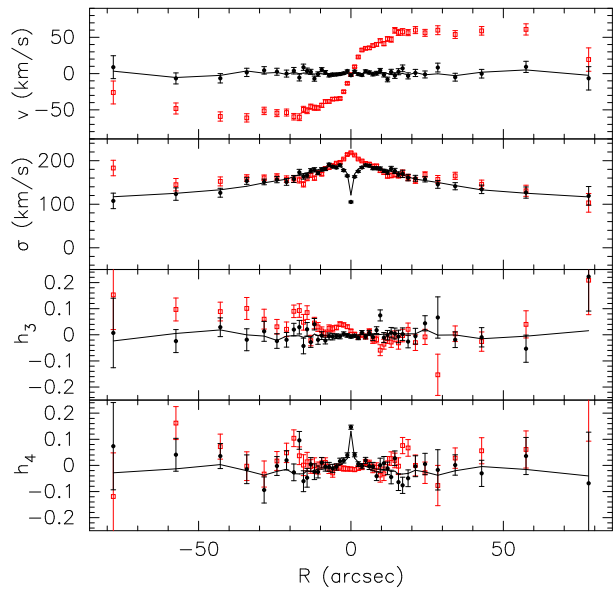}
\vskip0.2truecm
\caption[]{Comparison of the LOSVD kinematics of the mock galaxy model
  with those of NGC 3379 along its major axis, and with the particle
  model fit for $\mu=2\times10^3$.  The red open squares show the NGC
  3379 data from \citet{statler_smha99}, the black circles show the
  pseudo data, and the solid line the self-consistent
  particle model obtained from fitting the pseudo data.  The model
  data points are averages over the slit cells (see
  Fig.~\ref{fig:kinsetup}), and are connected by straight line
  segments.  The panels from top to bottom are for $v$, $\sigma$,
  $h_3$ and $h_4$.}
\label{fig:kin_mj_iso_vs_n3379}
\end{figure}

For the distribution function of the mock galaxy we take an
Osipkov-Merritt model \citep{osipkov79,merritt85} with anisotropy
radius $r_a=9a$, giving an anisotropy profile similar to some of our
later models for NGC 3379. The LOSVD kinematics is calculated
following \citet{carollo_etal95} and setting the mass-to-light ratio
to $\Upsilon_B=5$. To the final LOSVD parameters we add Gaussian
random variates with $1\sigma$ dispersions equal to the respective
error bars of the corresponding NGC 3379 measurements at that point.
In this way we compute $v$, $\sigma$, $h_3$ and $h_4$ points for the
mock galaxy along all slits shown in Figure \ref{fig:kinsetup}.
Figure \ref{fig:kin_mj_iso_vs_n3379} compares the kinematics of NGC
3379 with the mock galaxy model along the major axis.
%

\begin{figure*}
\centering
\includegraphics[width=0.6\hsize,angle=-90.0]{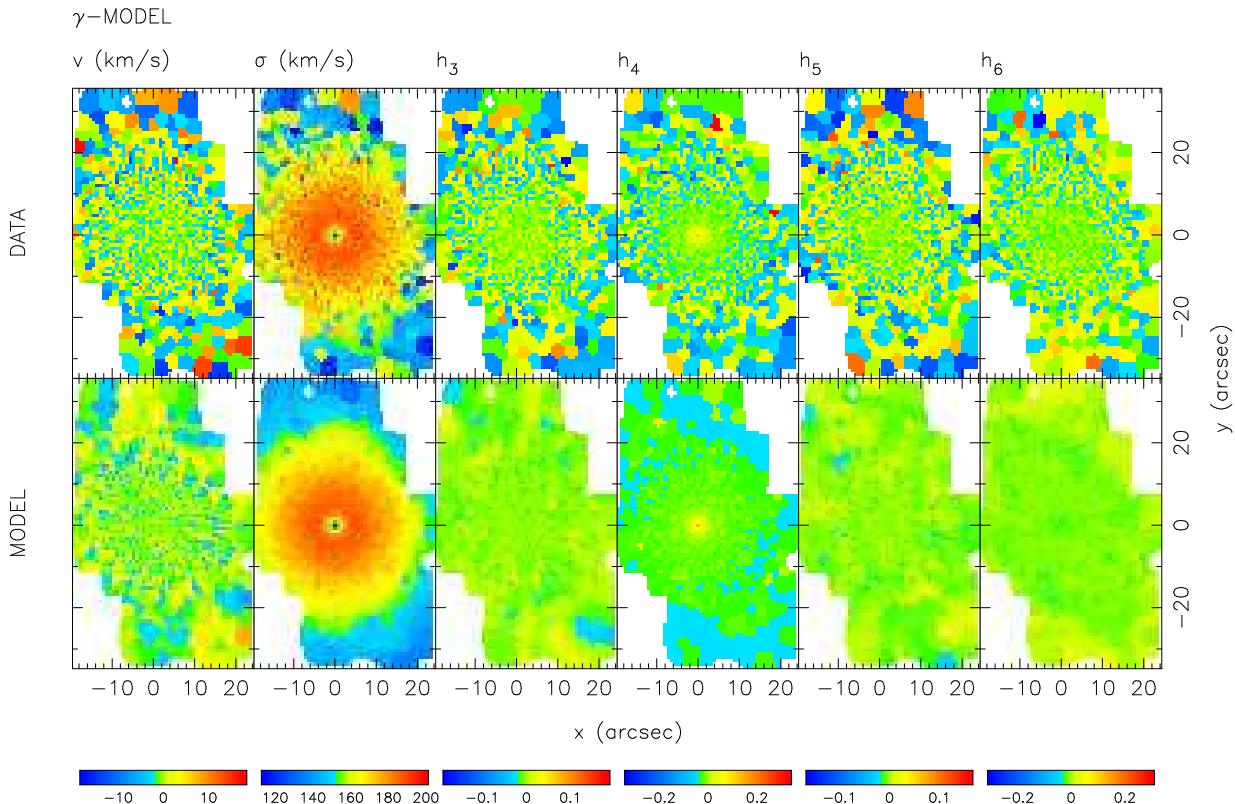}
\vskip0.5truecm
\vskip0.5truecm
\caption[]{Top panel: SAURON mock kinematic data for an anisotropic
  spherical galaxy model.  Bottom panel: Self-consistent particle
  realization obtained from a model fit with $\mu=2\times10^3$. From
  left to right: $v$, $\sigma$ and the higher order moments
  $h_3$-$h_6$.}
\label{fig:sauisomod}
\end{figure*}

In addition, we construct SAURON mock kinematics for each voronoi cell
in the NGC 3379 data as follows. We first compute the velocity
profiles as above at a few nearby radial positions.  Using the
spherical symmetry, we interpolate $v$, $\sigma$ and the higher order
moments to the mid-cell positions of the fine grid described in
Section \ref{sec:kindata}, using a spline interpolation scheme. Then
we compute the mock data for each voronoi bin by a luminosity weighted
average over those cells of the fine grid which contribute to the
voronoi cell under consideration. Finally, we add Gaussian random
variates to the kinematics with $1\sigma$ dispersions corresponding to
the respective SAURON error bars in this voronoi bin.  The SAURON
pseudo data are shown in the top panels of Figure \ref{fig:sauisomod}.
%
%
\begin{figure}
\centering
\includegraphics[width=0.9\hsize,angle=-90.0]{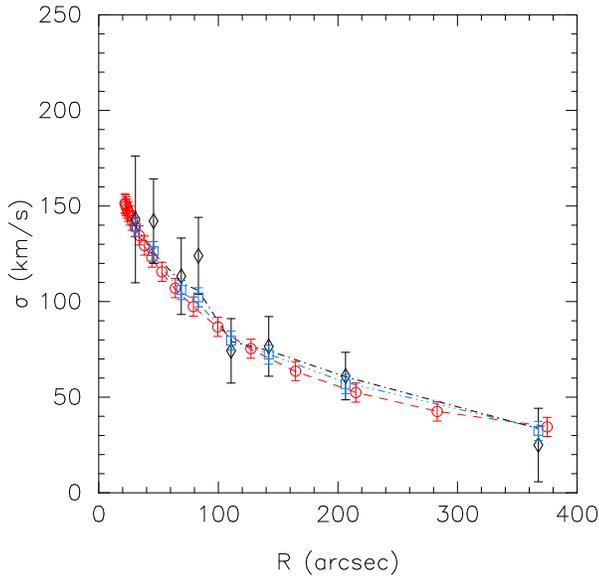}
\vskip0.2truecm
\caption[]{Different mock PN velocity dispersion data for the
    anisotropic spherical target galaxy model, and corresponding model
    velocity dispersion profiles obtained with $\mu=2\times10^3$. The
    three mock data sets differ by the quality with which they
    represent the underlying target velocity dispersion profile. The
    black diamonds have been obtained using the original errors of the
    \citet{douglas+07} PNe velocity dispersion data. The blue squares
    show the data obtained using reduced errors of $5 \kmsin$, and the
    red circles illustrate an idealized data set with an increased
    radial resolution. The corresponding NMAGIC models are presented
    by the black dash-dotted line, the blue dash-dot-dot-dot line and
    the red dashed line, respectively. The latter is indistinguishable
    from the target velocity dispersion profile.}
\label{fig:pnanisomod}
\end{figure}

The kinematic data set is completed with mock PN dispersion data,
using the projected velocity dispersions from \citet{carollo_etal95}
and the errors from the observed PN dispersion points in NGC 3379. For
test purposes, we also use two other sets of mock PN data with smaller
errors (see Figure~\ref{fig:pnanisomod}).  Again, Gaussian random
variates corresponding to these errors are added to account for the
scatter in the velocity dispersion points. Finally, we complete the
mock observational data set with the photometric constraints. In the
entropy tests here, we restrict ourselves to spherical models, so in
the expansion of the luminosity density the only non-zero term in the
spherical harmonics series (\cf Section \ref{sec:target}) is the
radial light in shells, $L_k=\sqrt{4\pi} A_{00,k}$. However, to ensure
sphericity, we also need to use the higher order coefficients
$A_{20,k}$, $\cdots$, $A_{22,k}$ and $A_{66,k}$ as constraints, set to
zero. We define these photometric observables on a grid of radii
$r_k$, quasi-logarithmically spaced in radius with inner and outer
boundaries at $r_{\rm min}=0.01\arcsec$ and $r_{\rm max}=2500\arcsec$.
We assume Poisson errors for the radial light $\sigma(L_k)=\sqrt{L_k
L/N}$ where N is the total number of particles used in the particle
model and $L$ is the total light of the system. To estimate the errors
in the higher order luminosity moments, we use Monte-Carlo experiments
in which we compute the $A_{lm}$ many times from random rotations of a
particle realization of the target density distribution. In these
experiments the number of particles is $7.5 \times 10^5$, which is the
same number as in the $\chi^2$M2M models.

We now construct self-consistent particle models for the
anisotropic model galaxy target in a two step process, using the mock
observations as constraints for NMAGIC. First, we start with the
particle model described in Section \ref{sec:fits} and evolve it using
NMAGIC to generate a self-consistent particle realization with the
desired luminosity distribution (mock particle model), fitting
only the photometric constraints. Then, we use the mock particle
model as initial conditions to fit both the kinematic and photometric
target constraints for different values of $\mu$.
%
\begin{figure}
\centering
\includegraphics[angle=-90.0,width=0.95\hsize]{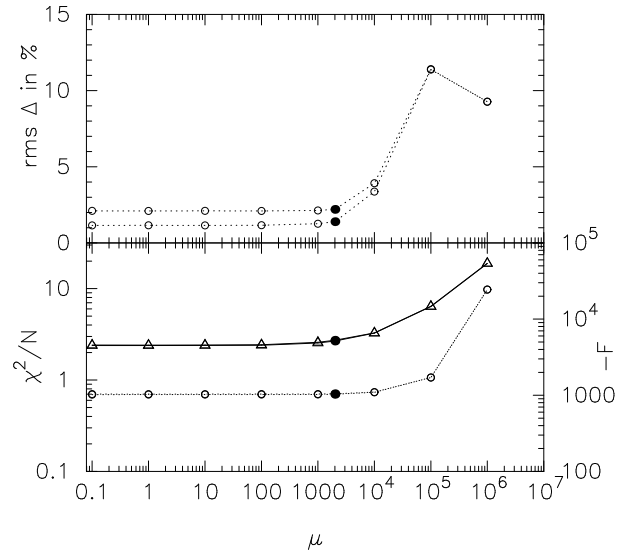}
\vskip0.2truecm
\caption[]{Top: Deviation rms $\Delta(\mu)$ between the internal
  velocity moments of the final mock galaxy particle model and the
  input model. The upper and lower curves show the rms $\Delta(\mu)$
  obtained, respectively, with the original mock PN velocity
  dispersion points, and with the errors of these dispersions reduced
  to $5 \kmsin$.  Bottom: The circles show $\chi^2$ per data point of
  the model fit to the kinematic and photometric targets as a function
  of entropy parameter $\mu$. The triangles display the merit function
  $F$, \cf equation (\ref{eqn:F}). For both quantities, the curves
  obtained with the two sets of errors fall on top of each other. The
  full symbol indicates the optimal value chosen for $\mu$.}
\label{fig:chimu}
\end{figure}

The results are presented in Figure \ref{fig:chimu}. The lower panel
shows the goodness of the fit as a function of $\mu$, both in terms of
the normalized $\chi^2$ per data point and in terms of the merit
function $F$ from equation (\ref{eqn:F}). The upper panel shows the
rms relative difference $\Delta$ between the true internal velocity
moments of the mock galaxy and those of the particle model
realizations obtained for different values of $\mu$.  The intrinsic
kinematics of the particle models are computed by binning the
particles in spherical polar coordinates, using a quasi-logarithmic
grid with $21$ radial shells bounded by $r_{\rm min} = 0.01 \arcsec$
and $r_{\rm max} = 500.0 \arcsec$, 12 bins in azimuthal angle $\phi$,
and 21 bins equally spaced in $\sin \theta$.  As can be seen from the
top panel of Figure \ref{fig:chimu}, there is no minimum in the rms
$\Delta$ as a function of $\mu$, but the particle models recover the
internal moments of the input model well for $\mu\apprla 2\times
10^{3}$.  For larger $\mu$, the rms $\Delta$ increases rapidly because
of oversmoothing in the model.  The lower panel of the figure shows
that $\chi^2$ per data point is below unity for a large range of $\mu$
but then increases for $\mu \apprga 2\times10^3$.  In our modelling of
NGC 3379 below we have confirmed that this value of $\mu$ allows the
models to converge towards strongly anisotropic orbit distributions.
We have therefore used $\mu=2\times 10^3$ in Section \ref{sec:models}
throughout. This is indicated by the solid symbol in
Fig.~\ref{fig:chimu}.

\begin{figure}
\centering
\includegraphics[angle=-90.0,width=0.95\hsize]{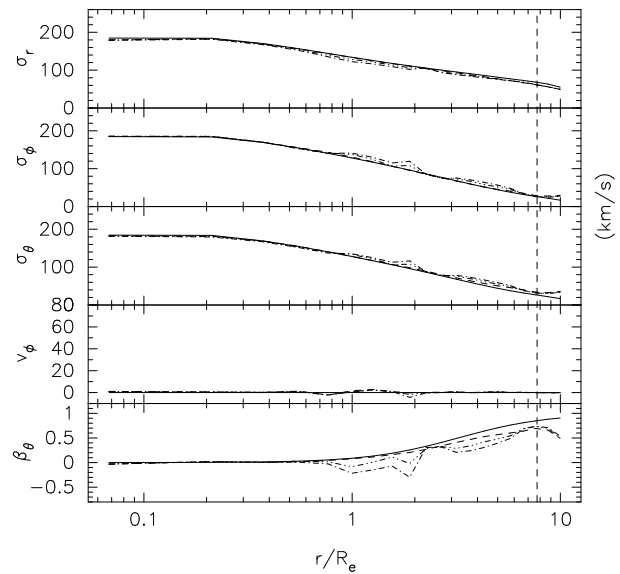}
\vskip0.2truecm
\caption[]{Internal kinematics of the anisotropic target galaxy and
  its final particle model realizations. From top to bottom: $\sigma_r$,
  $\sigma_\phi$, $\sigma_\theta$, $v_\phi$ and anisotropy parameter
  $\beta_\theta=1-\sigma_r^2/\sigma_\theta^2$. The kinematic
  quantities of the input mock galaxy are shown by circles, and are
  compared to those of three different particle models generated for
  $\mu=2\times10^3$: for idealized PN data (many points with small
  errors, dashed line), for mock data equivalent to the corresponding
  NGC 3379 data (dot-dashed), and to NGC 3379-like PN data with errors
  and scatter in the dispersion points reduced to $5\kmsin$; see
  Fig.~\ref{fig:pnanisomod}.}
\label{fig:intkin_iso}
\end{figure}
%
\begin{figure}
\centering
\includegraphics[angle=-90.0,width=0.95\hsize]{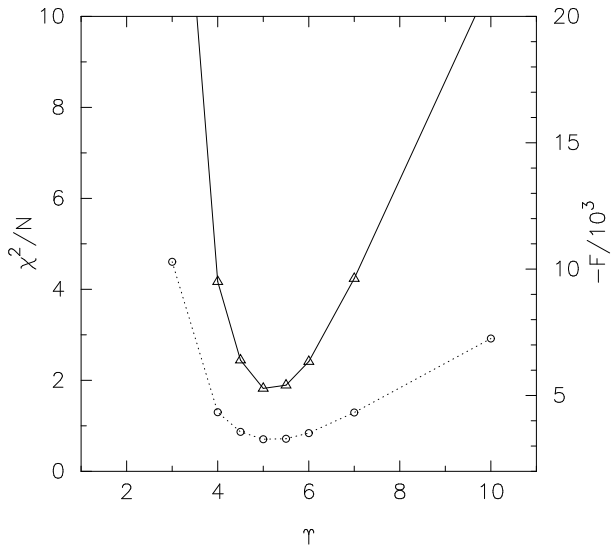}
\vskip0.2truecm
\caption[]{Recovering the mass-to-light ratio of the mock galaxy.  The
  quality of the model fit as a function of mass-to-light $\Upsilon$
  is shown in terms of $\chi^2$ per data point (circles) and merit
  function $F$ (triangles). All models are generated from the target
  pseudo data for $\mu=2\times 10^3$. The input mass-to-light ratio
  $\Upsilon=5$ is recovered as the minimum in the curve
  $\chi^2/N(\Upsilon)$, where the model fit has $\chi^2/N\simeq 0.7$.}
\label{fig:chim2l}
\end{figure}

Figures \ref{fig:kin_mj_iso_vs_n3379}, \ref{fig:sauisomod} and
\ref{fig:intkin_iso} compare the anisotropic mock galaxy model and 
the particle model obtained from the target data with
$\mu=2\times10^3$. Figure \ref{fig:kin_mj_iso_vs_n3379} shows the
target kinematics along the galactic major axis and the corresponding
particle model kinematics.  Figure \ref{fig:sauisomod} compares the
SAURON mock data with the two-dimensional kinematics obtained from the
particle model. Figure \ref{fig:pnanisomod} shows the PN velocity
dispersion data and compares to the model dispersion profiles. All
kinematic data are fit very well by the model.  In fact, it is evident
from Fig.~\ref{fig:sauisomod} that the model is smoother than the mock
data themselves, which is a consequence of entropy-smoothing and
time-smoothing.

Figure \ref{fig:intkin_iso} shows how well the internal kinematics of
the particle model for $\mu=2\times10^3$ compare with the intrinsic
kinematics of the mock galaxy target. The velocity dispersions
$\sigma_r$, $\sigma_\phi$ and $\sigma_\theta$, the streaming rotation
$v_\phi$, and also the anisotropy parameter
$\beta_\theta=1-\sigma^2_\theta/\sigma^2_r$ are reproduced well by the
model within $\simeq 1 R_e$.  At larger radii, the anisotropy of the
Osipkov-Merritt distribution function cannot be entirely recovered
even with idealized data (many dispersion points with small error
bars), because there is no constraint from the data on the outer
particle model (beyond the dashed line in
Fig.~\ref{fig:intkin_iso}). This is consistent with similar tests in
\citet{thomas+04}. With fewer mock PN dispersion data points and
larger errors, the particle model obtained from the target data and
isotropic initial conditions is less tightly constrained; it becomes
even less radially anisotropic despite fitting the actual PN data
points well.

\subparagraph*{Mass-to-light ratio}

So far all model fits have been made with the mass-to-light ratio
fixed to the actual value used for the mock galaxy, $\Upsilon=5$. Now
we investigate how accurately we can recover
$\Upsilon$ with the dynamical models, given the spatial extent and
quality of the observational data. To this end we fit particle models
to the mock galaxy observations for different mass-to-light ratios in
the range $\Upsilon \in [3,10]$, keeping the entropy parameter fixed
at $\mu=2\times 10^3$. The results are presented in Figure
\ref{fig:chim2l}, which shows how the quality of the model fit varies
as a function of $\Upsilon$, both in terms of $\chi^2$ per data point
and merit function $F$. As expected, the best model is obtained for
$\Upsilon=5$; it has $\chi^2$ per data point approximately 0.7.

\section{Dynamical models of NGC 3379}
\label{sec:models}
\begin{figure}
\centering
\includegraphics[angle=-90.0,width=0.95\hsize]{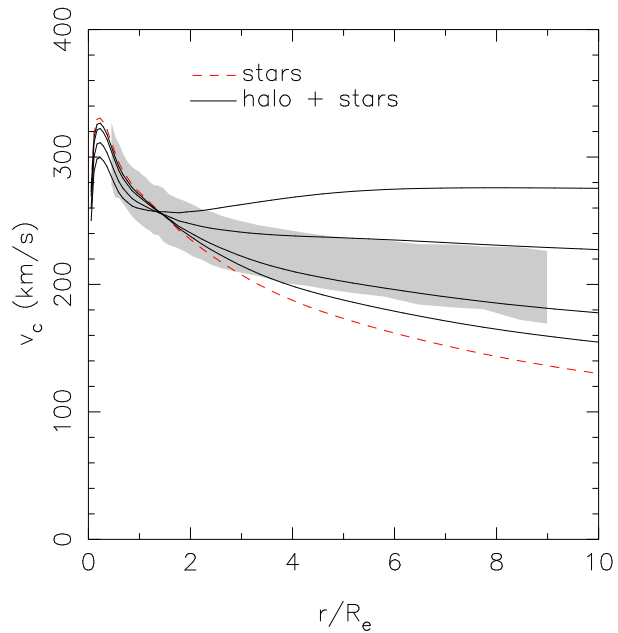}
\vskip0.2truecm
\caption[]{Circular velocity curves for the potentials used in the
  dynamical modelling, including the self-consistent stars-only model
  A (dashed line), and models including different spherical dark
  matter halos in addition to the stellar component (solid lines, from
  bottom to top: models B, C, D and E).  For this figure the
  distribution of stars is assumed to be spherical with mass-to-light
  ratio as given by the final NMAGIC model for the data in the respective
  spherical potential.  The shaded area shows the range of circular
  velocity curves in the merger models discussed by \citet{dekel+05};
  see Section~\ref{sec:conclusions}.}
\label{fig:circvel}
\end{figure}

In this section we construct dynamical models for NGC 3379 to learn
about its stellar and dark matter distribution. We investigate mainly
spherical and axisymmetric models, with and without dark matter halos,
and fit the photometry, SAURON integral field data, slit kinematics,
and PNe velocity data. Our aim in this paper is not to constrain the
detailed halo mass profile of the galaxy, but only to ascertain
whether a dark matter halo is allowed, or required, by the kinematic
data. Thus, as in \citet{delo+08} we investigate a simple sequence of
potentials which include the  self-consistent part from
the stellar component and a  fixed halo potential as in equation
(\ref{eqn:logpot}).  The circular speed curves corresponding to these
potentials vary at large radii from the near-Keplerian decline
expected when the mass in stars dominates, to the nearly flat shapes
generated by massive halos. They are shown in Figure \ref{fig:circvel}
and their halo potential parameters are given in Table
\ref{tab:haloparams}.
%

In the following subsections, we describe spherical models
(\S\ref{sec:sphmodels}) and oblate models (\S\ref{sec:oblatemodels}),
as well as a few models without imposed axisymmetry constraints
(\S\ref{sec:noaxi}), and then discuss the significance of the fits to
the data in a separate subsection (\S\ref{sec:fitquality}).  To begin
with we construct self-consistent particle models for NGC 3379 in
which the distribution of stars is spherical. This allows for an easy
comparison with previous work \citep{romanowsky_etal03,douglas+07}.
\subsection{Spherical models}
\label{sec:sphmodels}
\begin{table*}
\vbox{\hfil
\begin{tabular}{ccccccccccc}
  \hline \textsc{Halo} & $r_0/R_{e}$ & $v_0/km s^{-1}$ & $\chi^2/N$ &
  $\chi^2_{alm}/N_{alm}$ & $\chi^2_{sb}/N_{sb}$ &
  $\chi^2_{sau}/N_{sau}$ & $\chi^2_{sl}/N_{sl}$ & $\chi^2_{PN}/N_{PN}$
  & $-F$ & $\Upsilon$ \\ \hline

  A &  $0$ & $0$   & $0.208$ & $0.137$ & $-$ & $0.176$ &
  $0.565$ & $0.371$ & $2131.3$ & $8.23$\\

  B &  $3$ & $90$   & $0.215$ & $0.162$ & $-$ & $0.184$ &
  $0.548$ & $0.323$ & $2231.9$ & $8.03$\\

  C &  $3$ & $130$   & $0.216$ & $0.201$ & $-$ & $0.184$ &
  $0.539$ & $0.340$ & $2320.1$ & $7.82$\\

  D &  $3$ & $200$   & $0.219$ & $0.271$ & $-$ & $0.186$ &
  $0.522$ & $0.564$ & $2622.9$ & $7.28$\\

  D$^{+}$ &  $3$ & $200$   & $0.362$ & $0.641$ & $-$ & $0.300$ &
  $0.814$ & $1.002$ & $4409.2$ & $7.57$\\

  E &  $3$ & $260$   & $0.237$ & $0.484$ & $-$ & $0.192$ &
  $0.535$ & $1.557$ & $3175.2$ & $6.73$\\

  \hline

  E$^{*}$ &  $3$ & $260$   & $0.241$ & $-$ & $0.084$ & $0.215$ &
  $0.522$ & $0.504$ & $2649.4$ & $6.52$\\

  \hline
\end{tabular}
\hfil}
\caption{
  Table of parameters and fit results for models of NGC 3379 with
  spherical potentials. Models A-E correspond to the circular rotation 
  curves in Fig.~\ref{fig:circvel}. Model
  D$^+$ is the same as D but for a higher value of the entropy. 
  Model E$^\ast$ is the self-flattened oblate model in halo E of 
  Section \ref{sec:oblate_spot}. For these models columns (1)-(3)
  give the model code and the parameters $r_0$ and $v_0$ used in
  equation (\ref{eqn:logpot}) for the respective dark halo potential. 
  The next six columns list the $\chi^2$ values per
  data point, for all observables [column (4)] and for the
  luminosity density and surface brightness constraints, the SAURON
  kinematic observables, slit kinematic observables, and PN
  observables separately [columns (5)-(9)]. Column (10) gives the
  numerical value of the merit function in equation (\ref{eqn:F}), and
  column (11) the final (B-band) mass-to-light ratio. The respective
  number of constraints are $N=12997$ for A-E and $N=12557$ for
  E$^{*}$, whith $N_{alm}=640$, $N_{sb}=200$, $N_{sau}=11214$,
  $N_{sl}=1135$, $N_{PN}=8$.}
\label{tab:haloparams}
\end{table*}

\subsubsection{Target data and modeling process}

First we must determine the
photometric and kinematic observables. Analogous to Section
\ref{sec:isorot}, we use the spherical harmonics expansion
coefficients $A_{lm}$ of the deprojected luminosity density as target
data to constrain the particle models.  Specifically, we use $A_{00}$,
$A_{20}$, $A_{22}$, $\cdots$, $A_{66}$, but set all terms higher than
$A_{00}$ to zero, adopting the same radial grid as in Section
\ref{sec:isorot}.  Errors for the luminosity terms are estimated as in
Section \ref{sec:isorot}.  As kinematic observables, we use the SAURON
and slit kinematics, as well as the binned PN velocity dispersion
profile; see Sections \ref{sec:kindata} and \ref{sec:target}. The
SAURON data and most slit data are symmetrized, only the slit parallel
to their minor axis of \citet{krona_etal00} cannot be symmetrized and
for this slit the original kinematic data points are used.

We match the particle models to these data in the following three-step
process.  (i) We begin with the initial particle distribution described
in Section \ref{sec:fits} and evolve it with NMAGIC to a
self-consistent model that reproduces the target $A_{lm}$. (ii)
Starting with this density model we then construct dynamical models,
fitting the full set of photometric {\sl and} kinematic target
observables. If the potential includes a dark matter halo, we first
relax the density model for $1000$ steps in the total gravitational
potential (\cf Section \ref{sec:mass}), assuming a mass-to-light ratio
of 8, to make sure that the model is in approximate
equilibrium.  After this relaxation phase, we
evolve the particle system for $\sim 10^5$ NMAGIC correction steps
while applying the complete set of constraints. During the correction
phase the mass-to-light ratio $\Upsilon$ is adjusted in parallel,
using its own force-of-change equation as given in \citet{delo+08}.
After each correction step, the potential generated by the particles
is updated but the dark matter potential (if present) is constant in
time.  In this process the entropy parameter has value
$\mu=2\times 10^3$; cf.\  Sections~\ref{sec:isorot} and
\ref{sec:entropysmooth}.  (iii) In the final step, we keep the global
potential constant and evolve the system freely for another $5000$
steps, without changing the particle weights (phase-mixing). This
completes the modeling process. Thereafter we generally evolve the
model with all potential terms active for a further 10000 steps to
test its stability. For reference, 10000 correction steps in the
self-consistent potential correspond to $\approx 110$ circular
rotation periods at $R_e$, or 5.8 Gyr.

\begin{figure}
\centering
\includegraphics[angle=-90.0,width=0.95\hsize]{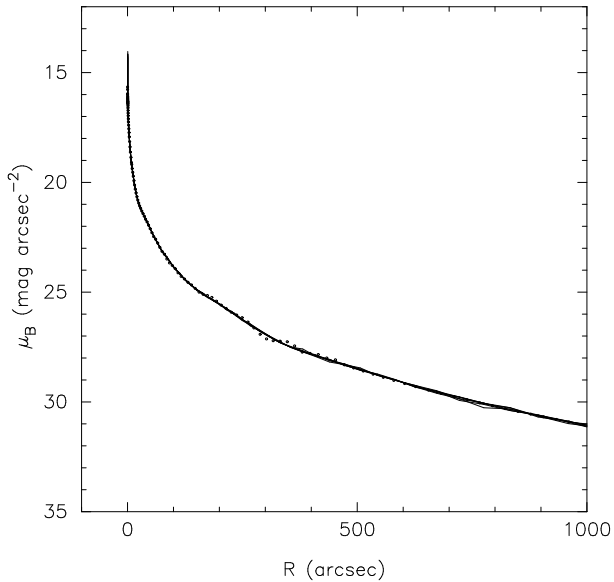}
\vskip0.2truecm
\caption[]{Comparison of the surface brightness profiles of the 
    reprojected spherical models with the photometric data (points).
  The lines are for the spherical models A-E and the self-flattened
  model E$^\ast$.}
\label{fig:sphSB}
\end{figure}

\begin{figure*}
\centering
\includegraphics[width=0.95\hsize,angle=-90.0]{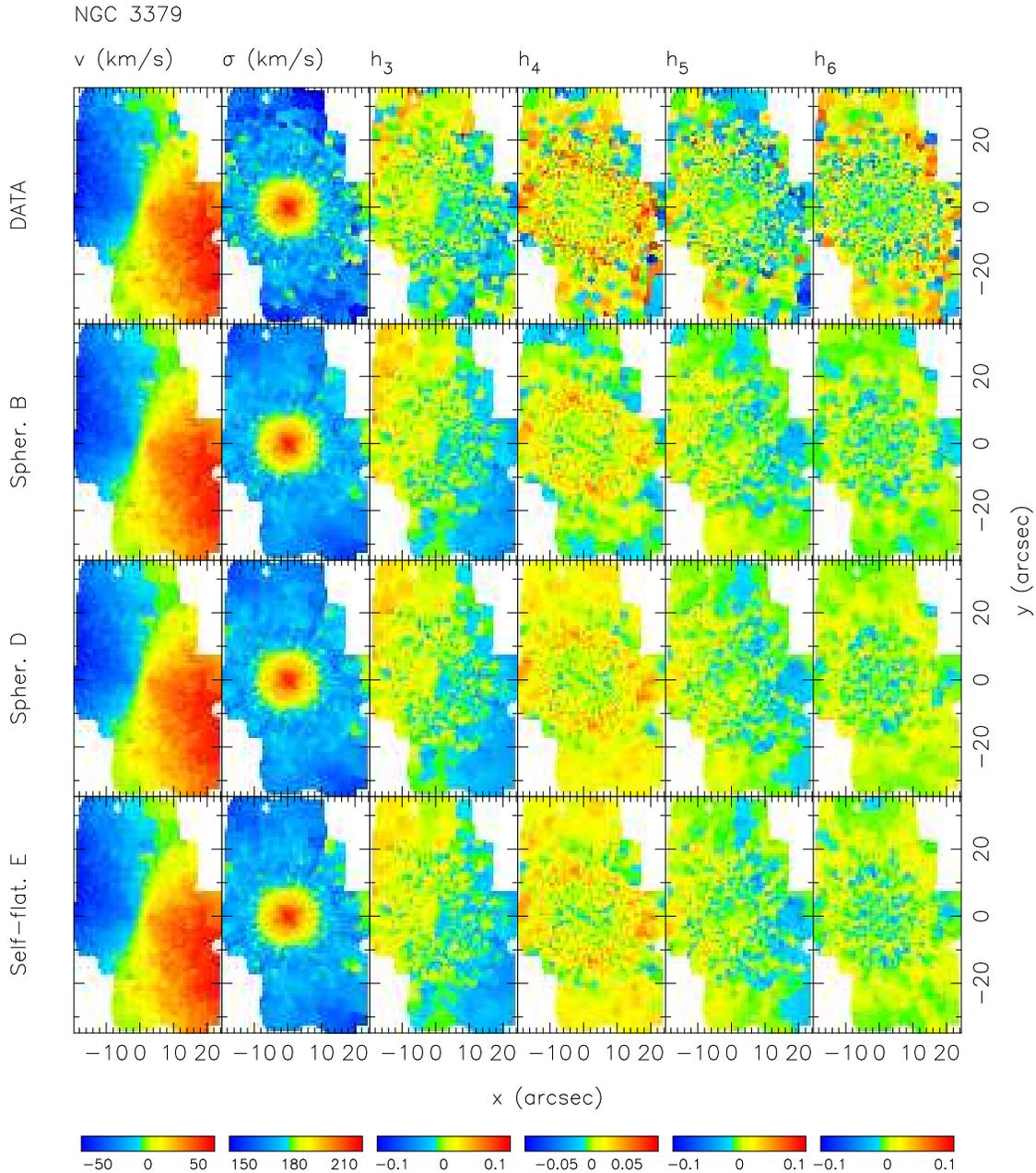}
\vskip0.5truecm
\vskip0.5truecm
\caption[]{Symmetrized SAURON kinematic data for NGC 3379 (top row)
  compared with similar data extracted for the spherical models B and
  D and the self-flattened model E$^\ast$ (lower three rows).  Notice
  that the particle noise in the model panels is significantly smaller
  than the noise in the corresponding data fields for all LOSVD
  parameters shown. In the panels for $\sigma$ and $h_4$ a slightly
  colder ring-like structure with larger $h_4$ hints at some
  deviations from spherical symmetry.}
\label{fig:sphsau}
\end{figure*}

\subsubsection{Results}

In this way we obtain spherical dynamical models for NGC 3379,
reproducing the density expansion and all kinematic data including the
PN velocity dispersion profile.  Model A is the self-consistent model
without dark matter halo, models B-E have halos of increasing circular
velocities, as shown in Figure \ref{fig:circvel} and Table
\ref{tab:haloparams}. The quality of fit for these models can be
judged from Table \ref{tab:haloparams}, which gives the numerical
values of the merit function $F$ and lists various values of $\chi^2$
per data point, both those obtained globally for all the data, and
those found for each of the four data sets separately (density
expansion, SAURON, slit, and PNe). The $3\sigma$ outlier point
discussed in Section~\ref{sec:pnedata} is not included in the modeling
and in the $\chi^2_{PN}$ in Table \ref{tab:haloparams}, but its
influence will be discussed below.

\begin{figure}
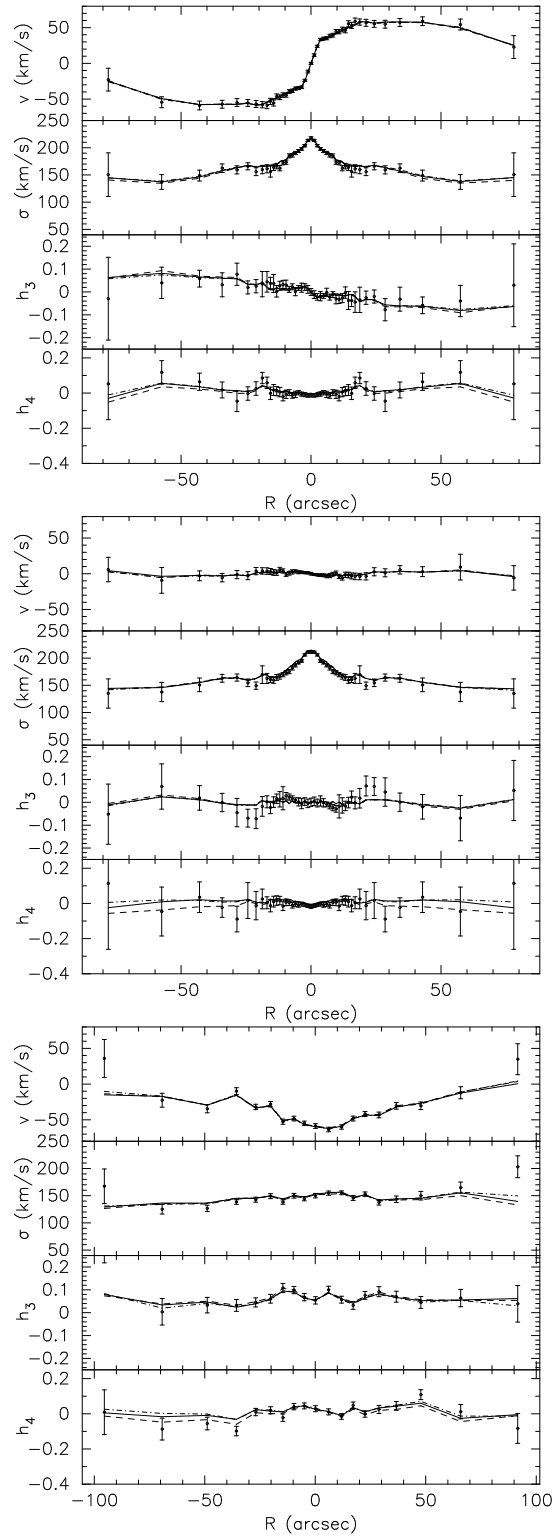

\centering
\includegraphics[angle=-90.0,width=0.85\hsize]{fig14a_sphStatl_mj.ps}
\includegraphics[angle=-90.0,width=0.85\hsize]{fig14b_sphStatl_mn.ps}
\includegraphics[angle=-90.0,width=0.85\hsize]{fig14f_sphKrona_para_mi.ps}
\vskip0.2truecm
\caption[]{Comparison of models B (dashed lines), D (full lines) and
  E$^\ast$ (dash-dotted lines) with the symmetrized slit data along
  the major and minor axes from \citet{statler_smha99} (top and middle
  panel) and the unsymmetrized minor-axis parallel slit from
  \citet{krona_etal00} (bottom panel). The model data points are
  averages over the same slit cells as the target data (see
  Fig.~\ref{fig:kinsetup}), and are connected by straight line
  segments.  }
\label{fig:sphslits}
\end{figure}

Figures \ref{fig:sphSB}-\ref{fig:sphpne} compare the different data
with the models.  Fig.~\ref{fig:sphSB} shows the surface brightness
profiles, Fig.~\ref{fig:sphsau} the integral field LOSVD parameter
fields, Fig.~\ref{fig:sphslits} the kinematics along several slits,
and Fig.~\ref{fig:sphpne} the PN velocity dispersion profiles.  The
model SB profiles fit the observed profile very well, and agree with
each other within the thickness of the lines in the plot. The SAURON
data are fitted with $\chi^2_{sau}/N_{sau}\simeq 0.2$ by all our
spherical models. Notice that the particle noise in the models is
significantly smaller than the noise in the symmetrized Sauron maps.
Also the $\chi^2_{sl}/N_{sl}$ for the combined slit data are less than
unity; the plots for models (B,D) in Fig.~\ref{fig:sphslits} show a
few small systematic deviations but generally the fits are very good.
In the central 30 arcsec the slit data are dominated by the SAURON
data.  Notice that these spherical models are not constrained to be
spherically symmetric also in their {\sl kinematic} properties; hence
they can also fit the  observed ($\apprla 50\kmsin$) rotation of NGC
3379 with high accuracy.   The small $\chi^2$-values are caused by
the fact that the observational errors are slightly larger than the
point-to-point fluctuations (see Fig.~\ref{fig:krona_vs_sau}), and to
a greater extent, because with the entropy scheme we cannot smooth the
models too much without erasing their anisotropic phase-space
structure; cf.\ Sections \ref{sec:isorot}, \ref{sec:entropysmooth} and
\citet{delo+08}. 

The comparison of the models to the PN.S data is shown in
Fig.~\ref{fig:sphpne}.  If we use the outermost dispersion point as
given in \citet{douglas+07}, models A-D with no or moderately massive
halos provide a good match to the data, but the most massive halo
model E fits less well, being high by $\simeq2\sigma$ with respect to
the outermost dispersion point and by $\simeq1.3\sigma$ with respect
to the second-outermost point. If we include the object classified as
$3\sigma$ ``friendless'' outlier \citep[see Section
\ref{sec:pnedata} and ][]{douglas+07} in the outermost bin,  the 
$1\sigma$ error range of the outermost PN dispersion point extends to
significantly larger velocities; see the red open circle and error
bar in Fig.~\ref{fig:sphpne}. Then model E also fits the PN dispersion
profile, overestimating the outermost velocity dispersion point by
less than $1\sigma$.

The intrinsic kinematics of these spherical models is shown in Figure
\ref{fig:sphintkin}. One recognizes the expected signature of the
well-known mass-anisotropy degeneracy \citep{bin+mamon82}: In the more
massive halos, the same falling line-of-sight dispersion profile
requires larger radial anisotropy. Thus in the models with halo the
radial anisotropy rises outside 1-2$R_e$. Particularly the more
massive halo models D and E require strongly radially anisotropic
orbit distributions ($\beta\simeq 0.9$) to be consistent with the
falling dispersion profile of NGC 3379.  Radial anisotropy was
suggested as one of the possible causes for the measured profile by
\citet{dekel+05}, based on a comparison with their merger models.
However, the typical anisotropies in their models are more moderate
($\beta\simeq 0.5$). 

Despite their strong radial anisotropy, the massive halo models D and
E show no sign of an instability when evolved freely after the model
fitting and phase-mixing.  Rather, they evolve very slowly, reaching
after 5.8 Gyr of evolution a configuration with slightly triaxial
shape ($\epsilon<0.1$) in which the initial slow rotation has mostly
gone away.  A similar evolution is seen for the near-isotropic model A
without dark matter halo, indicating that this evolution may be
connected to these equlibria being spherically symmetric only in their
mass distribution but, due to the rotation. not in their kinematics.
In any case, the PN dispersion profiles do not change during the
evolution, i.e., the constraints on the dark matter halo remain as
before.

In conclusion, the results of this section show that both
near-isotropic spherical models with low density dark matter halos,
and radially anisotropic spherical models with massive halos provide
excellent fits to the available kinematic data for NGC 3379, including
the PN dispersion profile to $\sim 7R_e$. A more quantitative
discussion is deferred to Section \ref{sec:fitquality}.

\subsubsection{Entropy smoothing}
\label{sec:entropysmooth}

\begin{figure}
\centering
\includegraphics[angle=-90.0,width=0.95\hsize]{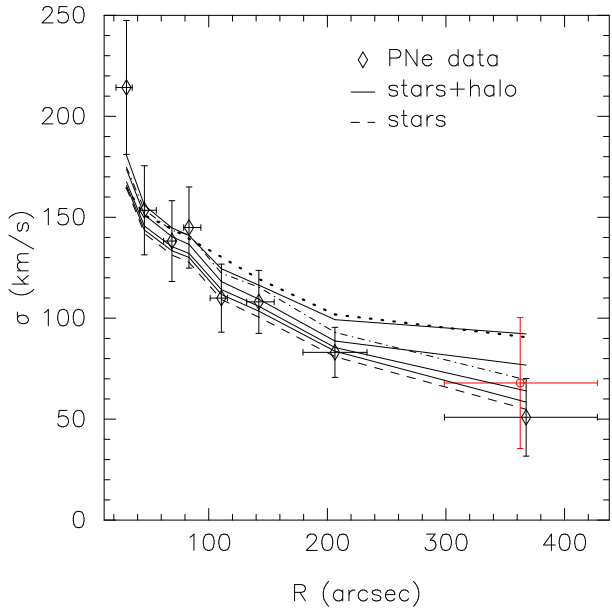}
\vskip0.2truecm
\caption[]{Comparison of the PNe velocity dispersion profiles of the
  spherical models with the PN.S data. The PNe velocity dispersion
  points of \citet{douglas+07} are shown as black diamonds; when the
  object classified by them as $3\sigma$ ``friendless'' outlier is
  included,  the $1\sigma$ error range of the outermost PN
  dispersion point extends to significantly larger velocities (red
  open circle). The dashed line shows the self-consistent particle
  model A.  The solid lines represent the dynamical models including a
  DM halo, i.e., from bottom to top models B, C, D, and E. The heavy
  dotted line is for the higher-entropy model D$^+$, and the
  dash-dotted line is for the self-flattened model E$^\ast$.}
\label{fig:sphpne}
\end{figure}

The entropy term in the force-of-change equation~(\ref{eqn:myFOC})
smoothes the particle models by trying to maintain the values of the
particle weights near their priors, here chosen as $1/N$. Because all
models start from an isotropic system with equal weight particles, the
entropy smoothing thus biases the final models towards isotropy and
slow rotation. To allow the models to develop strong radial anisotropy
in their outer parts  requires a relatively low value of the
entropy parameter (see Section \ref{sec:isorot}), which is below that
appropriate for an isotropic system. Otherwise the constraints from
the small number of PN dispersion points with their relatively large
Poisson error bars would be overwhelmed by the entropy smoothing.  We
demonstrate this in Fig.~\ref{fig:sphpne} and Table
\ref{tab:haloparams} with a model D$^{+}$ constructed with
$\mu=2\times 10^4$; this model is indeed degraded in its ability to
fit the PNe data, relative to model D which is for the standard
$\mu=2\times10^3$ in the same halo.  

\begin{figure}
\centering
\includegraphics[angle=-90.0,width=0.95\hsize]{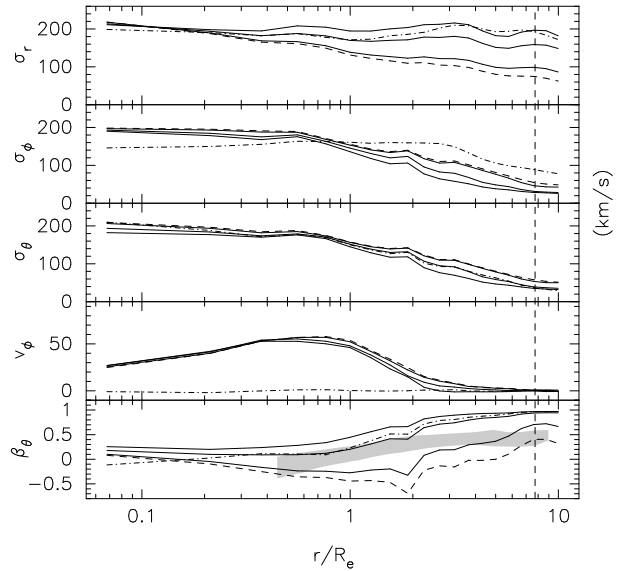}
\vskip0.2truecm
\caption[]{Intrinsic kinematics of the final spherical models A
  (dashed lines) and B,D,E (full lines), and the self-flattened model
  E$^\ast$ (dash-dotted lines). Panels from top to bottom show the
  radial, azimuthal, and vertical velocity dispersion profiles, the
  mean azimuthal streaming velocity, and the meridional anisotropy
  profile, all computed in an equatorial plane through the model
  (perpendicular to the rotation axis for the spherical models, and
  perpendicular to the line-of-sight for model E$^\ast$). The models
  in the more massive dark matter halos are more radially anisotropic,
  as expected. The shaded area in the lower panel corresponds to the
  range of anisotropy profiles found in the elliptical galaxy remnants
  in the merger simulations of \citet{dekel+05}.}
\label{fig:sphintkin}
\end{figure}

Contrary to second derivative regularisation, say, entropy smoothing
does not distinguish between local and global uniformity of the
particle weights; it likes to have {\sl all} particle weights similar
to their priors. Thus if $\mu$ is chosen such as to allow large
differences in weight between radial and circular orbits, it also
allows similar differences between particles on neighbouring orbits if
this is preferred by the data. With $\mu=2\times10^3$ the models can
therefore fit the data with $\chi^2/N< 1$ as seen in
Table~\ref{tab:haloparams}. The effect is strongest for the spherical
models because these have a larger number of independent orbits than
less symmetric systems.  However, Fig.~\ref{fig:sphintkin} shows that
the intrinsic velocity moments are smooth functions of radius, and
below we will see that also the LOSVDs are smooth functions. Thus the
good fits of the various models to the PN data are not achieved by
large local variations of the orbital weights for orbits near the PN
data points.

\subsection{Oblate models including dark matter halos}
\label{sec:oblatemodels}
There is some evidence that NGC 3379 may be non-spherical.
\cite{capaccioli+91} argued that the bulge of NGC 3379 is remarkably
similar to the one of NGC 3115, a well-known S0 galaxy. Further, also
the SAURON kinematic data, shown in the upper panel of Figure
\ref{fig:sphsau}, show signatures of non-sphericity, particularly, a
faint cold ring visible in the velocity dispersion and $h_4$ panels
with projected radius $R\approx 15\arcsec$\footnote{As can be seen
  from Fig.~\ref{fig:sphsau}, the feature can also be reproduced in
  spherical models.}.
Thus to understand how much dark mass around NGC 3379 is allowed by
the kinematic data for this galaxy may require more general models
than spherical ones. In this section we will present oblate
axisymmetric models in the family of  spherical halo potentials
considered already in the last section.

\begin{figure}
\centering
\includegraphics[angle=-90.0,width=0.95\hsize]{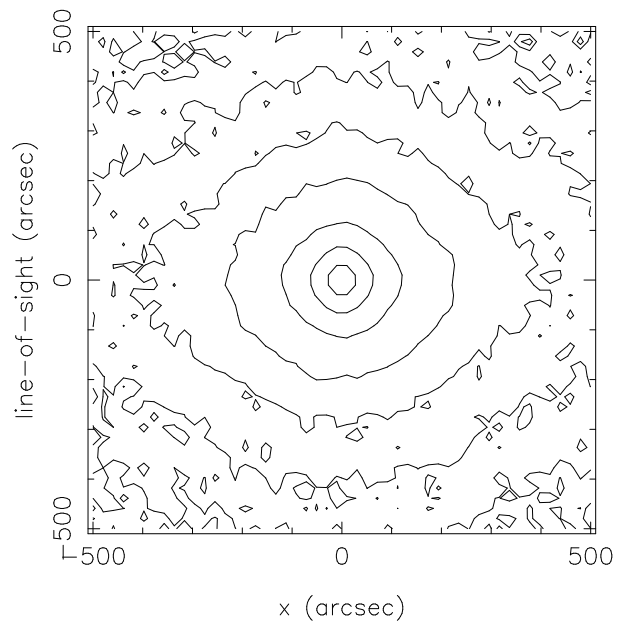}
\vskip0.2truecm
\caption[]{Surface brightness contours of the kinematically
  deprojected oblate model E$^\ast$, when viewed perpendicular to the
  line-of-sight, i.e., edge-on.  The model's outer parts have been
    preferentially flattened to match the falling PN velocity
    dispersion profile.}
\label{fig:oblSBcont}
\end{figure}

\subsubsection{Face-on oblate model in a  spherical potential}
\label{sec:oblate_spot}
As a first step we attempt to construct a model for NGC 3379 in a
massive dark halo, in which the distribution of stars is flattened
along the line-of-sight.  This model is required to have a small
line-of-sight velocity dispersion at large radii, thus will be
flattened in accordance with the virial theorem
\citep[\eg][]{bin_tre87}.  We do not know beforehand what the required
shape of this model must be, so we will use the NMAGIC method to find
it for us.  Throughout this experiment the gravitational potential
is constrained to remain spherically symmetric, being the sum of the
spherical part of the luminous matter potential and the spherical halo
potential. For illustration we embed this model in halo E, and will
hence hereafter denote it as model E$^{\ast}$.

To construct this model we replace the $A_{lm}$ constraints (\cf
Section \ref{sec:sphmodels}), which before imposed a spherical shape
on the particle distribution, by the Fourier moments of the surface
brightness distribution given in Figure \ref{fig:photo}. They are
computed from the photometry as in Section \ref{sec:target}, on a grid
in projected radius quasi-logarithmically spaced betwen $R_{\rm
  min}=0.01\arcsec$ and $R_{\rm max}=1500\arcsec$.  The higher-order
moments are set to zero, enforcing axisymmetry.  We then start from
spherical initial conditions and use NMAGIC to flatten the particle
model through fitting the kinematic observables, particularly the PN
velocity dispersion profile. As kinematic constraints, we use the
SAURON, slit, and PNe velocity dispersion data.  The entropy parameter
is kept at the same value as for the spherical models, $\mu=2\times
10^3$.  During this ``kinematic deprojection'', the spherically
averaged potential generated by the particles is updated after regular
time intervals, but the non-spherical terms are ignored. The DM
potential is given by equation (\ref{eqn:logpot}) and remains constant
in time.  After the correction phase, the model is again allowed to
freely evolve for some time.

\begin{table*}
\vbox{\hfil
\begin{tabular}{ccccccccccccc}
  \hline
  \textsc{Halo} & $r_0/R_{e}$ & $v_0/km s^{-1}$ &
   $i$ & $\chi^2/N$ & $\chi^2_{alm}/N_{alm}$ &
   $\chi^2_{sb}/N_{sb}$ & $\chi^2_{sau}/N_{sau}$ &
   $\chi^2_{sl}/N_{sl}$ & $\chi^2_{PN}/N_{PN}$ & $-F$
   & $\Upsilon$ \\
  \hline

  A90 &  $0$ & $0$   & $90$ & $0.619$ & $0.173$ & $0.331$ & $0.624$ &
  $0.866$ & $0.369$ & $4899.995$ & $8.09$\\

  A50 &  $0$ & $0$   & $50$ & $0.773$ & $0.291$ & $0.437$ & $0.781$ &
  $1.031$ & $0.426$ & $6129.759$ & $8.12$\\

  A40 &  $0$ & $0$   & $40$ & $0.789$ & $0.507$ & $0.587$ & $0.780$ &
  $1.079$ & $0.515$ & $6634.361$ & $8.22$\\

  \hline

  B90 &  $3$ & $90$   & $90$ & $0.631$ & $0.243$ & $0.40$ & $0.635$ &
  $0.852$ & $0.344$ & $5051.231$ & $7.923$\\

  B50 &  $3$ & $90$  & $50$ & $0.777$ & $0.352$ & $0.523$ & $0.782$ &
  $1.008$ & $0.371$ & $6196.835$ & $7.97$\\

  B40 &  $3$ & $90$  & $40$ & $0.782$ & $0.570$ & $0.670$ & $0.770$ &
  $1.047$ & $0.438$ & $6562.924$ & $8.10$\\

  \hline

  C90 &  $3$ & $130$  & $90$ & $0.651$ & $0.296$ & $0.457$ & $0.655$ &
  $0.851$ & $0.401$ & $5291.926$ & $7.72$\\

  C50 &  $3$ & $130$  & $50$ & $0.741$ & $0.429$ & $0.611$ & $0.742$ &
  $0.933$ & $0.396$ & $6030.503$ & $7.82$\\

  C40 &  $3$ & $130$  & $40$ & $0.766$ & $0.661$ & $0.591$ & $0.753$ &
  $0.990$ & $0.414$ & $6478.453$ & $7.98$\\

  \hline

  D90 &  $3$ & $200$  & $90$ & $0.611$ & $0.367$ & $0.462$ & $0.603$ &
  $0.847$ & $0.887$ & $5343.843$ & $7.26$\\

  D50 &  $3$ & $200$  & $50$ & $0.761$ & $0.394$ & $0.663$ & $0.763$ &
  $0.961$ & $0.815$ & $6394.686$ & $7.50$\\

  D40 &  $3$ & $200$  & $40$ & $0.745$ & $0.618$ & $0.639$ & $0.738$ &
  $0.906$ & $0.654$ & $6466.793$ & $7.693$\\

  \hline

  E90 &  $3$ & $260$  & $90$ & $0.684$ & $0.577$ & $0.751$ & $0.652$ &
  $1.037$ & $2.602$ & $6325.564$ & $6.86$\\

  E50 &  $3$ & $260$  & $50$ & $0.765$ & $0.530$ & $0.854$ & $0.749$ &
  $1.026$ & $2.401$ & $6782.397$ & $7.20$\\

  E40 &  $3$ & $260$  & $40$ & $0.756$ & $0.819$ & $0.739$ & $0.899$ &
  $0.737$ & $1.662$ & $6806.086$ & $7.42$\\

  \hline

  DR &  $3$ & $200$  & $50$ & $0.715$ & $-$ & $0.567$ & $0.699$ &
  $0.897$ & $0.890$ & $5990.4$ & $7.57$\\

  ER &  $3$ & $260$  & $50$ & $0.710$ & $-$ & $1.313$ & $0.676$ &
  $0.894$ & $6.317$ & $6219.4$ & $6.85$\\

  \hline

\end{tabular}
\hfil}
\caption{
  Table of parameters and $\chi^2$-fit results for oblate models of NGC 
  3379. Columns (1)-(3) give the model code and the parameters $r_0$, $v_0$ 
  used in equation (\ref{eqn:logpot}) for the respective dark halo
  potential; all halo potentials are spherical ($q_\phi=1.0)$. 
  The fourth column gives the inclination $i$
  and the next six columns list the $\chi^2$ values per data point,
  for all observables [column (5)], and for the density constraints,
  surface brightness constraints, SAURON kinematic observables, slit
  kinematic observables, and PN observables separately
  [columns (6)-(10)]. Column (11) gives the numerical value of the merit
  function in equation (\ref{eqn:F}), and column (12) the final (B-band)
  mass-to-light ratio. The respective number of constraints are
  $N=13237$, $N_{alm}=680$, $N_{sb}=200$, $N_{sau}=11214$,
  $N_{sl}=1135$, $N_{PN}=8$.}
\label{tab:oblate_result}
\end{table*}

Figures \ref{fig:sphSB}, \ref{fig:sphslits}, \ref{fig:sphsau} and
\ref{fig:sphpne} show how the final ``self-flattened'' particle model
E$^\ast$ compares to the various data. The model fits the data as well
as the best-fitting spherical models.  As anticipated, the model makes
the PN dispersion profile compatible with a massive dark halo
potential by flattening the outer distribution of stars and decreasing
the model $\sigma$ along the line-of-sight.  Fig.~\ref{fig:sphintkin}
shows that the line-of-sight velocity dispersion measured at $\sim
2-7R_e$ in the equatorial plane ($\sigma_\theta$) is half the
$\phi$-dispersion in this plane; the radial dispersion still
dominates, however.  The model's flattening is illustrated in Figure
\ref{fig:oblSBcont}, which shows the SB distribution in an edge-on
projection perpendicular to the line-of-sight. The axis ratio is
$q\simeq 0.7$, but to match the decreasing line-of-sight velocity
dispersion profile, the flattening increases at large radii.

While this model illustrates the power of the NMAGIC method, and
provides an excellent fit to the photometric and kinematic data in a
massive dark DM halo, it is not a realistic model for NGC 3379. For it
is only in a spherical potential as assumed for model E$^\ast$ that a
face-on distribution of stars can show rotation. More realistic
axisymmetric models must therefore be inclined to allow for the
rotation seen in the SAURON and slit data.

\subsubsection{Self-consistent oblate models}
\label{sec:oblate}

\begin{figure*}
\centering
\includegraphics[width=0.95\hsize,angle=-90.0]{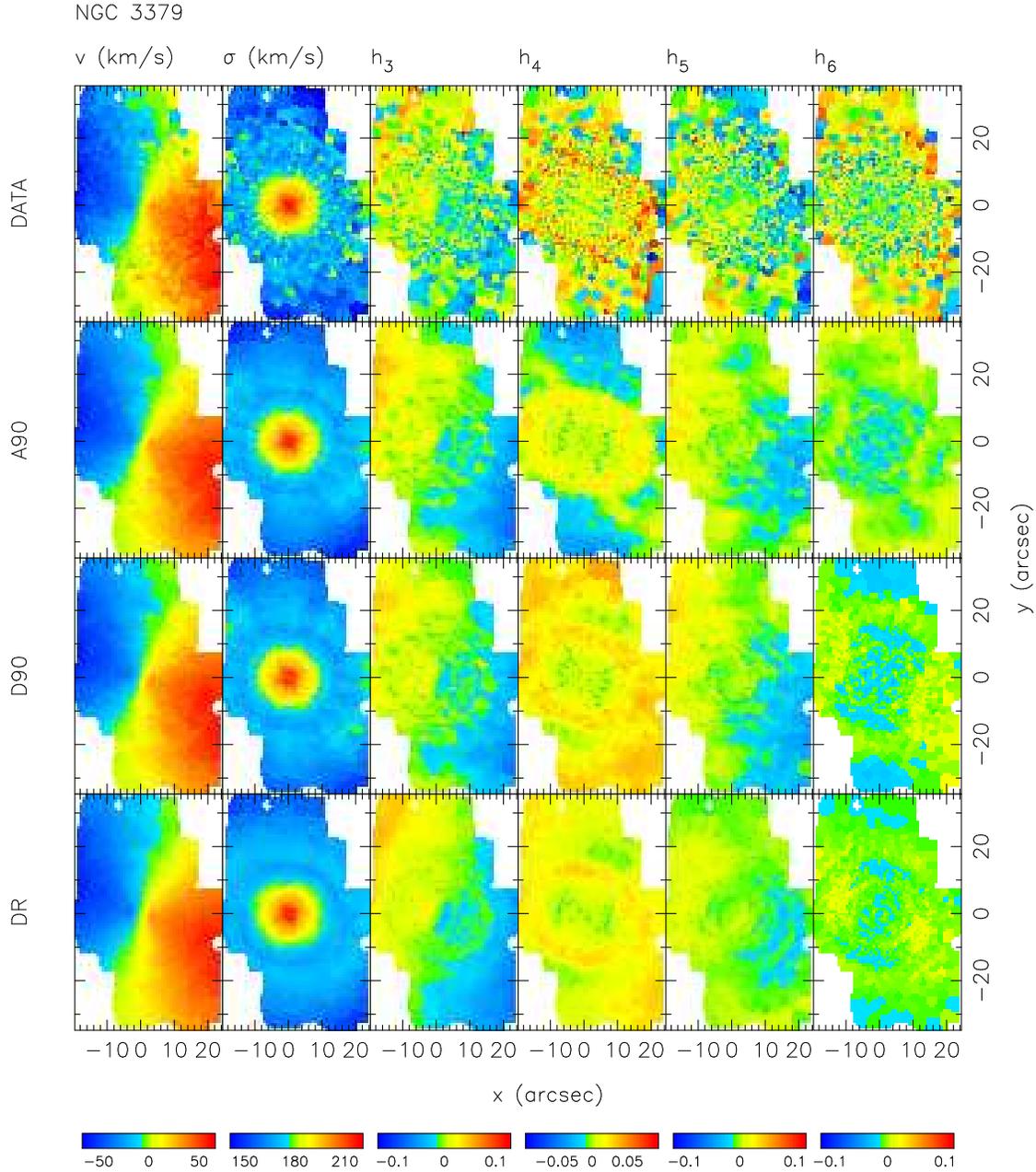}
\vskip0.5truecm
\vskip0.5truecm
\caption[]{Comparison of axisymmetric and weakly triaxial models with 
  SAURON kinematic data for NGC 3379 (top panel). Following panels are
  for models A90, D90, DR. Model A90 has all the mass in the stars,
  while the later two models include a massive halo; see Table
  \ref{tab:oblate_result}.}
\label{fig:axisau}
\end{figure*}

Therefore we now consider oblate models for NGC 3379 with
inclinations $i=90\degrees$, $i=50\degrees$ and $i=40\degrees$, in
which the axisymmetric gravitational potential of the stellar
component is computed self-consistently from the particles.  We
investigate models without DM as well as models including various DM
halos as detailed in Table \ref{tab:oblate_result}.  The gravitational
potential of the DM halo is still assumed to be spherical and is kept
fixed.  All models are evolved to fit the photometry, the slit and SAURON
absorption line kinematic data, and the PN.S velocity dispersion
profile.

The procedure employed for constructing the models is similar to that
in Section \ref{sec:sphmodels}. Again, we need to first specify the
observables. We expand the deprojected luminosity density of NGC 3379
for each inclination in a spherical harmonics series and determine the
expansion coefficients on the same quasi-logarithmic grid in radius as
before. As observables we use the luminosity on radial shells $L_k$
and the higher order moments $A_{20}$, $A_{22}$, $\cdots$, $A_{66}$,
but set the $m\ne0$ terms of the expansion to zero to force the models
to remain axisymmetric.  Errors for the $A_{lm}$ coefficients are
estimated as in Section \ref{sec:isorot}. We thus obtain three
different sets of luminosity density observables $A_{lm}$ with
corresponding errors, one for each of the three inclinations. In
addition to the $A_{lm}$, we also use the surface brightness itself
as a constraint, through
the Fourier moment observables on the grid of projected radii
$R_k$ as in the previous Section \ref{sec:oblate_spot}. Errors for
these Fourier moments are computed similarly as the $A_{lm}$ errors.
The kinematic constraints are identical to those used for the
spherical models: they are the luminosity weighted, symmetrized
Gauss-Hermite moments from the slit data and SAURON data (see Section
\ref{sec:kindata}), and the PNe kinematics represented by the binned
line-of-sight velocity dispersion points.

For the combined set of observables we construct particle models in a
similar three-step process as for the spherical models. (i) We start
with the spherical particle model described in Section
\ref{sec:target} and use NMAGIC to generate an equilibrium model with
the desired luminosity distribution, as given by the deprojection of
the photometry for the given inclination. (ii) We then use the
resulting particle model as a starting point to generate the final set
of models by fitting the photometric and kinematic constraints in the
different DM halos.  We use the same entropy parameter
$\mu=2\times10^3$ as for the spherical models.  (iii) Finally, we
first keep the potential constant and let the system evolve freely
without changing the particle weights, and thereafter test the
stability of the model.

The quality of the fit for the different halo models and inclinations
is again characterized by the value of the merit function $F$ of
equation (\ref{eqn:F}) and the values of the different $\chi^2$ per
data point, both globally and for the individual data sets. These are
given in Table \ref{tab:oblate_result} and will be discussed further
in Section \ref{sec:fitquality}. In addition to the models shown in
Table \ref{tab:oblate_result}, we have also constructed a similar
suite of models for the unsymmetrized SAURON and slit data. These
models were of similar quality as the models for the symmetrized data,
i.e., when subtracting the systematic error floors determined in
Section \ref{sec:absndata} ($\chi^2_{sau}/N_{sau}({\rm sys})=1.0$ and
$\chi^2_{sl}/N_{sl}({\rm sys})=1.0$) from the $\chi^2$ values of the
models for the unsymmetrized data, the model $\chi^2$ values became
very similar to those reported in Table \ref{tab:oblate_result}.

\begin{figure}
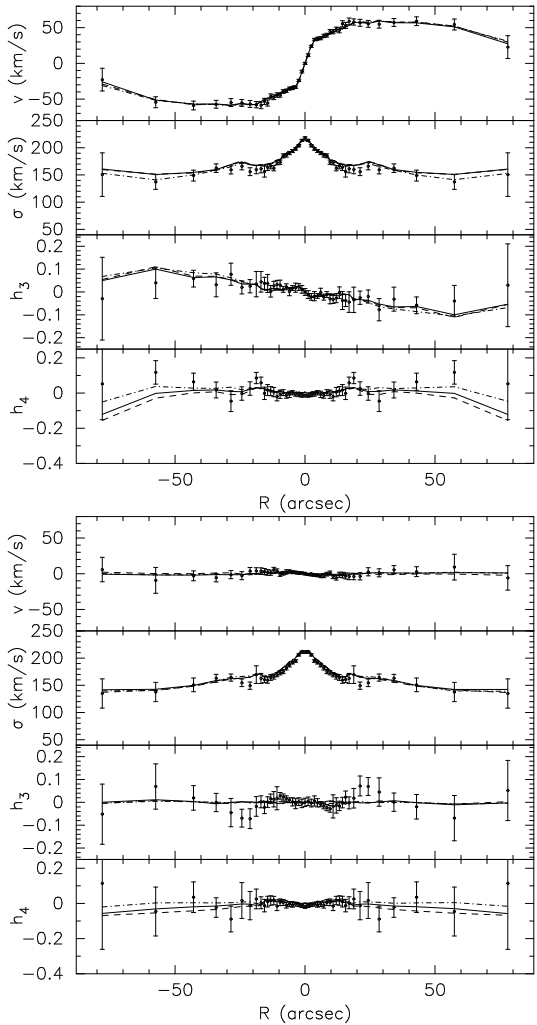

\centering
\includegraphics[angle=-90.0,width=0.85\hsize]{fig19a_oblStatl_mj_DR.ps}
\includegraphics[angle=-90.0,width=0.85\hsize]{fig19b_oblStatl_mn_DR.ps}
\vskip0.2truecm
\caption[]{Comparison of the axisymmetric models A90 (dashed lines), D90 
  (full lines) and the weakly triaxial model DR (dash-dotted lines)
  with the symmetrized slit data from \citet{statler_smha99} along the
  major (top) and minor axes (bottom panel). The model data points are
  averages over the same slit cells as the target data (see
  Fig.~\ref{fig:kinsetup}), and are connected by straight line
  segments.  }
\label{fig:axislits}
\end{figure}

\begin{figure}
\centering
\includegraphics[angle=-90.0,width=0.95\hsize]{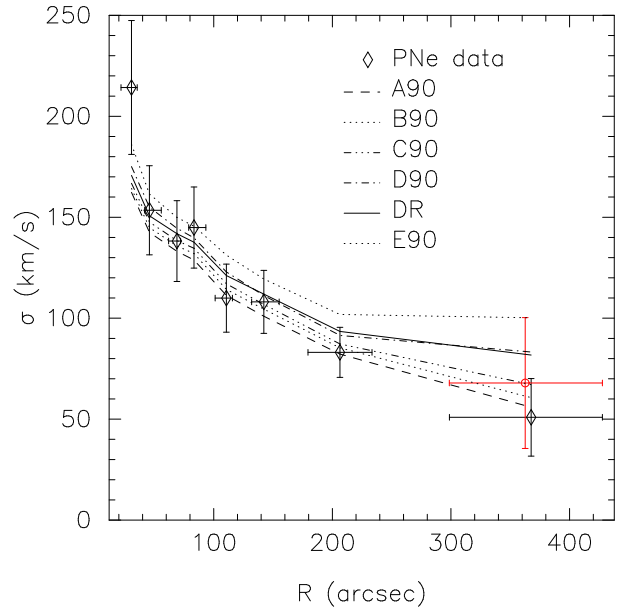}
\vskip0.2truecm
\caption[]{Comparison of the radial velocity dispersion profile from
  the PN.S data with the oblate and weakly triaxial particle models. The
  dashed line shows the stellar-mass only model A90. The other broken
  lines show models B90, C90, D90, the solid line shows model DR, and
  the upper dotted line shows model E90.}
\label{fig:axipne}
\end{figure}

\begin{figure}
\centering
\includegraphics[angle=-90.0,width=0.95\hsize]{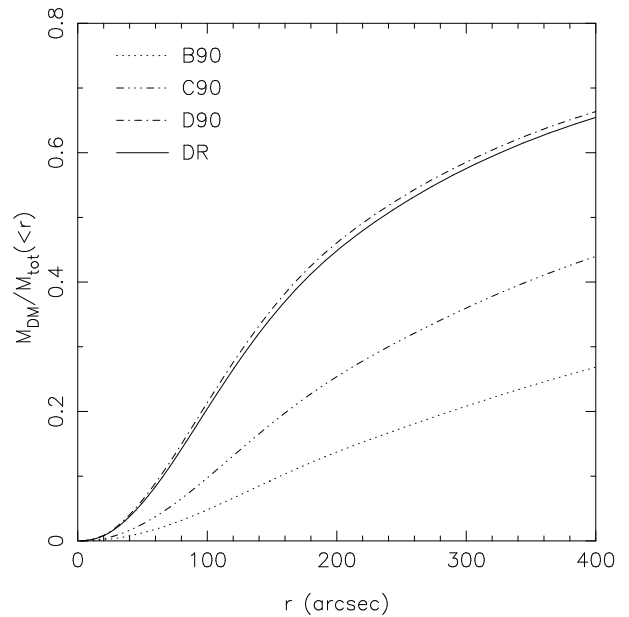}
\vskip0.2truecm
\caption[]{Enclosed DM fraction as function of radius for the final
  particle models B90, C90, D90, DR.}
\label{fig:DMfrac}
\end{figure}

Figures \ref{fig:axisau}-\ref{fig:axipne} compare some of the final
axisymmetric particle models to the SAURON, slit and PNe data. Both
edge-on and inclined models again are very good matches to the SAURON
and slit data, with or without dark matter halo.  The PN velocity
dispersion profile is fitted well by the models with the lower mass
halo models B,C; halo D slightly overestimates the outer PN velocity
dispersion point given by \citet{douglas+07} but is within $1\sigma$ of
the outer point when the ``friendless'' outlier is included.
Model E90 is inconsistent with the outer dispersion point of
\citet{douglas+07}, cf.\ Table~\ref{tab:oblate_result}, but
is only marginally inconsistent with the data when the outlier is
included. Based on this together with the likelihood results reported
below, halo D is the most massive halo consistent with the PN data.
Figure \ref{fig:DMfrac} shows that for this model the dark halo
contributes about 60\% of the total mass within the radius of the last
PN data point at $\sim 7R_e \sim 15\kpc$.

%
%
%
\begin{figure}
\centering
\includegraphics[angle=-90.0,width=0.95\hsize]{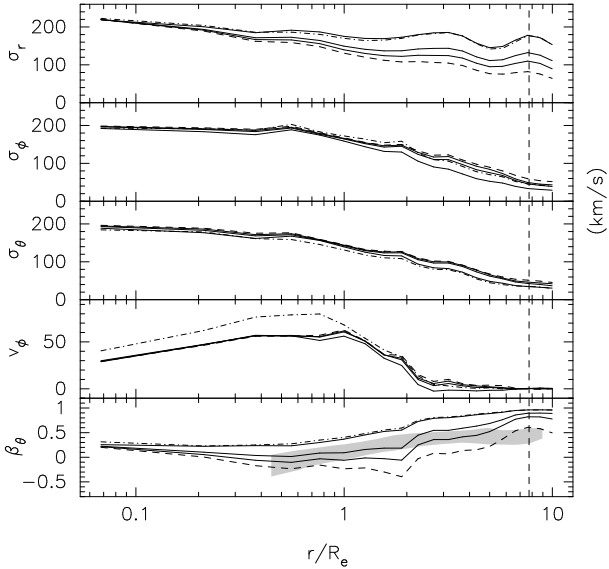}
\vskip0.2truecm
\caption[]{Intrinsic kinematics of the final models A90, B90 (dashed),
  C90, D90 (full) and D50 (dash-dotted lines). Panels from top to
  bottom show the radial, azimuthal, and $\theta$ velocity dispersion
  profiles, the mean azimuthal streaming velocity, and the meridional
  anisotropy profile $\beta_\theta=1-\sigma_r^2/\sigma_\theta^2$. 
  The models in the most massive halos are
  strongly radially anisotropic, as expected.}
\label{fig:axiintkin}
\end{figure}

Figure \ref{fig:axiintkin} shows the intrinsic velocity dispersions,
streaming velocity, and anisotropy for some of the models. Because of
the small projected ellipticity of NGC 3379,  and the assumed
spherical dark matter halo, the edge-on models are
very similar to the spherical models in the respective halo potentials
and the higher circular velocity halos require large radial anisotropy
to match the PN data. The inclined flattened models have similarly
small $\sigma_z=\sigma_\theta$ in the model equatorial plane, but
somewhat larger $\sigma_\phi$, as expected. Also in the axisymmetric
models it is the radially increasing, strong radial anisotropy which
causes the rapidly decreasing PN velocity dispersion profile in the
massive dark halo potentials.

Finally, we comment briefly on the stability of these models.  All
models in halos A-C show no signs of any change after 5.8 Gyr of
evolution following the phase-mixing after the NMAGIC fit. The D
models are almost unchanged, despite the strong radial anisotropy,
developing after 5.8 Gyr a percent-level triaxiality just outside the
error bars of the $A_{22}$ constraints.  The models in halo E show a
similar slow evolution during which they in addition develop
significant positive $h_4$ across the entire image. As in the
spherical models, the PN dispersion profiles remain unchanged during
this evolution.

\subsection{Triaxial models}
\label{sec:noaxi}

We have constructed a small number of models for which the stellar
density was not constrained to remain axisymmetric, in order to see
whether the larger freedom in the orbit structure of non-axisymmetric
potentials would allow the models to fit the PN kinematics also in the
most massive halo E. However, these models do not have isophote
twists: we have kept the constant value PA$=70\deg$ for the position
angle in the photometry, neglecting the observed small variations
$\Delta{\rm PA}=\pm3\deg$. These models are generated as follows,
using the full power of NMAGIC: Only the surface brightness and
kinematic data are used as constraints, in a similar way as for model
E$^\ast$, leaving all density $A_{lm}$ terms and corresponding
potential terms free to change during the evolution. This allows the
model to freely change its orientation. As initial conditions we have
used a spherical model, a model flattened along the line-of-sight, or
the inclined model D40.

Because we know that valid models in halo D can be found, we first
evolve a model in halo D, starting from initial conditions D$^\ast$, a
model that had previously been obtained exactly analogously to model
E$^\ast$ (see Section~\ref{sec:oblate_spot}).  Because of the
line-of-sight streaming velocities, this system rotates out of the sky
plane while NMAGIC simultaneously keeps adjusting the orbit structure
to match both the surface brightness and the projected kinematics.
 This model converges to an almost axisymmetric model with inclination
{$i\simeq 46\deg$}, and is then completely stable over 5.8 Gyr of
evolution. We have computed iteratively the mean intermediate and
minor axis lengths inside ellipsoidal radius
$s=[x^2+(y/b)^2+(z/c)^2]^{1/2}$, following
\citet{dubinski_carlberg91}. We obtain axis ratios $b=0.988$ and
$c=0.74$ for $q = 50 \arcsec \simeq 2.375 \kpc$, and $b=0.990$ and
$c=0.73$ for $q = 200 \arcsec \simeq 9.5 \kpc$. Note that the error in
these axis ratios is about $0.002$, due to the large number of particles
used in the diagonalization of the tensor.  This weakly triaxial model
matches all the kinematic data, SAURON, slit, and PNe, very similar to
models D40 and D50, and is listed in Table~\ref{tab:oblate_result} as
model DR. Its projected kinematics are shown in
Figs.~\ref{fig:axisau}-\ref{fig:axipne}, and the enclosed dark matter
fraction is about 60\% of the total mass within the radius of the last
PN data point at $\sim 7R_e \sim 15\kpc$ (Fig.~\ref{fig:DMfrac}).

Also shown in the Table are the results for model ER, which was
obtained analogously starting from model E$^\ast$. This model does not
fit the PN data. None of our other attempts to obtain a valid model E
has been successful, including one inspired by some old work on merger
remnants \citep{gerhard83a,gerhard83b}, following which we tried to
construct an  oblate-triaxial model whose inner oblate parts are
  seen edge-on by the observer, while its triaxial outer regions are
  observed along the short axis. 

We believe the main reason for the failure in halo E is the observed
rotation of NGC 3379, of which either the sense (along the projected
major axis) or the amplitude do not allow the low-inclination
configurations required by the low values of velocity dispersion at
large radii.  Consider a triaxial model viewed approximately along the
short axis, which could easily accomodate the outer falling dispersion
profile by a corresponding change of shape with radius as in model
E$^\ast$.  Such a model is not consistent with the observed rotation
because in this case the rotation visible to the observer would be
around the model's long axis, i.e., it would be observed along the
minor axis on the sky, whereas the actual observed rotation is along
the major axis on the sky. On the other hand, a triaxial model with
minor axis in the sky plane must correspond to a very round model
unless it is near-prolate and viewed end-on. Then the box orbits are
also viewed near end-on, making it difficult to arrange a falling
outer dispersion profile. Thus valid triaxial models are likely to be
radially anisotropic and to have inclinations near $i\simeq45\deg$ as
model DR, or be more edge-on and quite round, so the dominant effect
on the inferred potential is the radial anisotropy.  If so, the
failure of model ER then suggests that extending our suite of models
to include triaxial halos (beyond the scope of this paper), is
unlikely to increase the range of allowed circular velocities.

\subsection{Likelihoods and quality of the fits to the data}
\label{sec:fitquality}

We now turn to discussing the question which models are acceptable
fits to the data and which models can be ruled out. To do this, it is
customary to determine $\Delta\chi^2$ values relative to the
best-fitting models, and determine the confidence boundaries according
to the number of parameters to be determined. In our case, we
essentially determine only one parameter, the halo circular velocity
at $\sim 7R_e$, or $v_0$, so the relevant $\Delta\chi^2=1$ (the
mass-to-light ratio of the models is optimized together with the
weights).  However, all our models match the Sauron and slit kinematic
data to within $1\sigma$ per data point, i.e.,  formally better
than the underlying ``true'' model  (cf.\ the discussion in
  Sections \ref{sec:isorot}, \ref{sec:entropysmooth} and
  \citet{delo+08}).  Clearly, we cannot apply a $\Delta\chi^2=1$ for
small variations within $1\sigma$ relative to, say, the Sauron data
points.  Even if the best model fitted with exactly
$\chi^2_{sau}\simeq 10^4$, this would make little sense: for
$N_{sau}=10^4$, $\Delta\chi^2=1$ corresponds to an average change per
data point of $\simeq 10^{-4}\sigma$. Only if the $\Delta\chi^2=1$
arises because of significant mismatch of a few crucial data points
would this seem reasonable. The crucial data points for the issue
addressed in this paper, the dark matter halo in NGC 3379, are the PN
velocities or the binned PN dispersions. Thus we focus our discussion
on the merit of the models relative to these data.

\begin{figure}
\centering
\includegraphics[angle=-90.0,width=0.95\hsize]{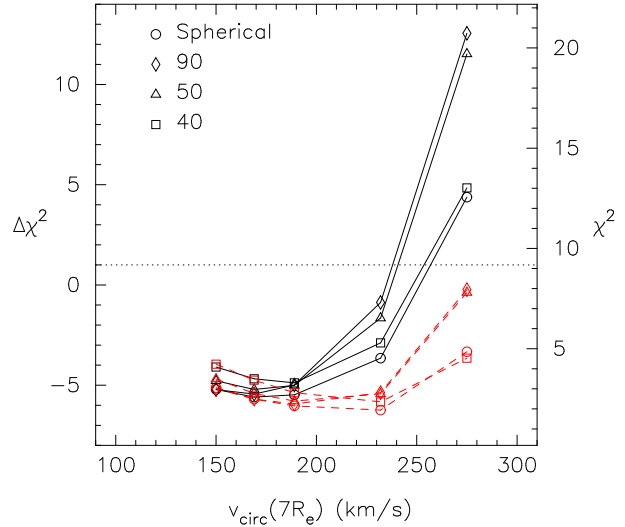}
\vskip0.2truecm
\caption[]{$\chi^2$ and $\Delta\chi^2$ values of the various spherical
  and axisymmetric models for NGC 3379 with respect to the PN velocity
  dispersion data. The full black lines (red dashed lines) connect the
  $\chi^2$-values obtained without (including) the $3\sigma$
  "friendless" outlier of \citet{douglas+07}.  $\Delta\chi^2$ is
  computed relative to the expected value of $\chi^2=8.18$ for 7
  degrees of freedom.}
\label{fig:allchisq}
\end{figure}

Figure \ref{fig:allchisq} shows the $\chi_{PN}^2$ and $\Delta\chi^2$
values for both the spherical and the axisymmetric models from Tables
\ref{tab:haloparams} and \ref{tab:oblate_result}. For the PN
dispersion points we have 7 degrees of freedom (8 data points minus 1
fitted parameter), so expect $\chi^2=8.18$ (68.3\% probability) for a
typical good model. Thus we consider any model that fits the PN
velocity dispersions to better than $\chi^2=8.18$ as valid as the
underlying ``true'' model and compute $\Delta\chi^2$ relative to
$\chi^2_{PN}=8.18$. The curves in Figure \ref{fig:allchisq} are
plotted for the two cases with and without the ``friendless'' outlier
of \citet{douglas+07} contributing to the outermost dispersion point.
The models with halos A-D are allowed in both cases, while the models for
halo E are consistent with the data only when the outlier is included.

\begin{figure}
\centering
\includegraphics[angle=-90.0,width=0.95\hsize]{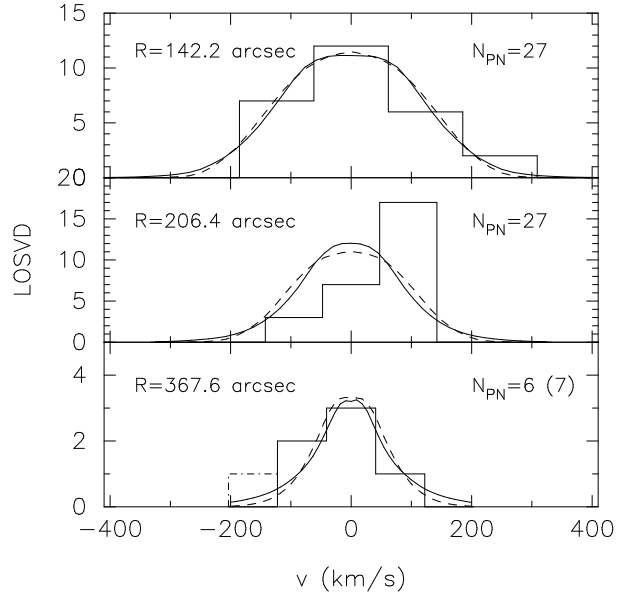}
\vskip0.2truecm
\caption[]{Comparison of the PNe LOSVDs in the circular annuli
  corresponding to the outermost three bins in the velocity dispersion
  profile, with the LOSVDs of the nearly isotropic, low-density halo
  model B  (dashed lines), and the radially anisotropic massive
  halo model D  (solid lines), in the same circular annuli. The
  ordinate is in units of PN number, and the model LOSVDs have been
  normalized to the same integral over the velocity range shown. In
  the middle panel for the second annulus, the mean velocity of the
  PNe is non-zero at the $\simeq 3\sigma$ level; both models are
  inconsistent with this velocity distribution.  The velocity
  distributions in the other two panels are fully consistent with both
  models. In the lower panel for the outermost shell, the PN histogram
  is shown with (dash-dotted) and without (solid line histogram) the
  $3\sigma$-'"friendless"' outlier; see Section \ref{sec:pnedata} and
  \citet{douglas+07}.  }
\label{fig:sphlosvd}
\end{figure}

So far we have compared the models only to the PN velocity dispersion
profile, rather than to the LOSVDs or unbinned velocity data.  Figure
\ref{fig:sphlosvd} shows the LOSVD histograms for the PNe in the
outermost three circular annuli used for computing the PN velocity
dispersion profile, superposed on the LOSVDs of models B and D in the
same radial shells. In the plot for the outermost bin, the PN
histogram and model LOSVD are shown with and without the $3\sigma$
``friendless'' outlier according to \citet{douglas+07}. Both the
near-isotropic low-density halo model B and the radially anisotropic
massive halo model D are consistent with the PN velocity distributions
in the first and third annuli, and both appear inconsistent with the
 apparent non-zero mean motion of the  observed PNe in the
second annulus.

\begin{table}
\vbox{\hfil
\begin{tabular}{ccccc}
  \hline  & \multicolumn{2}{c}{without outlier} &
  \multicolumn{2}{c}{with outlier} \\

  \hline \textsc{Halo} & $\ln{\mathcal L}$ & $2\Delta\ln{\mathcal L}$ &
  $\ln{\mathcal L}$ & $2\Delta\ln{\mathcal L}$ \\ \hline
  A & $-605.14$ & $2.50$ & $-611.18$ &	$4.58$ \\
  B & $-604.21$ & $0.64$ & $-609.67$ & $1.56$\\		
  C & $-603.89$ & $0.0$ & $-608.89$ & $0.0$\\		
  D & $-604.74$ & $1.7$ & $-609.23$ & $0.68$\\
  E & $-607.16$ & $6.54$ & $-611.38$ & $4.98$\\

  \hline
  A & $-608.50$ & $4.23$ & $-613.53$ & $5.01$ \\
  B & $-607.01$ & $1.25$ & $-611.82$ & $1.60$\\
  C & $-606.38$ & $0.0$  & $-611.02$ & $0.0$\\
  D & $-606.68$ & $0.60$ & $-611.14$ & $0.23$\\	
  E & $-608.81$ & $4.85$ & $-613.04$ & $4.02$\\

  \hline
\end{tabular}
\hfil}
\caption{Likelihood values for the PN data in the spherical models.
  Column (1): model code. Columns (2,3): log likelihood $\ln{\mathcal
    L}$ and difference $2\Delta\ln{\mathcal L}$ relative to the best
  model C, for the PN sample not including the $3\sigma$ "friendless"
  outlier in the outermost shell, according to
  \citet{douglas+07}. Columns (4,5): same, but for the PN sample
  including this outlier. The top half of the table refers to
  posterior likelihoods of the models fitted to the PN velocity
  dispersion profile, the lower half gives likelihoods for similar
  models in which the PNe were fitted with the likelhood method of
  \citet{delo+08}.}
\label{tab:lh}
\end{table}

\begin{figure}
\centering
\includegraphics[angle=0.0,width=0.95\hsize]{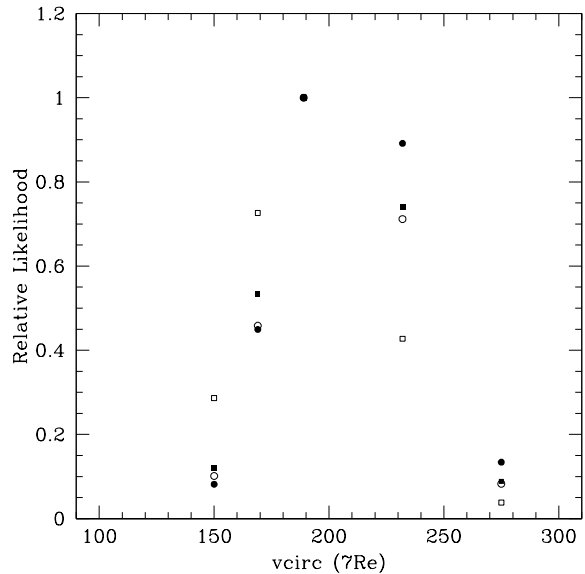}
\vskip0.2truecm
\caption[]{Relative likelihoods from the data in Table \ref{tab:lh} as
  a function of the model circular velocity at $7R_e$. Open symbols
  show posterior likelihoods of models fitted to the binned dispersion
  profile, full symbols show likelihood values based on direct
  likelihood fits to the PN velocities. Squares show likelihoods for
  the PN sample without the $3\sigma$ "friendless" outlier in the
  outermost shell, according to \citet{douglas+07}, circles for the
  sample including this outlier.}
\label{fig:lh}
\end{figure}

Table \ref{tab:lh} shows the posterior likelihoods of the spherical
models for the observed PN velocity data set, evaluated from the model
LOSVDs in the eight radial shells used in the fits. Also listed are
the likelihoods resulting from direct likelihood fits of the spherical
models to the PN data, using the method described in \citet{delo+08}.
Figure~\ref{fig:lh} shows a plot of these likelihoods as a function of
the models' circular velocity at $7R_e$, the radius of the outermost
PN dispersion point. Despite the small number of potentials
investigated and the issue of whether the $3\sigma$ "friendless"
outlier should be included, the overall shape of the likelihood
function ${\cal L}$  is not too far from 
the theoretically expected Gaussian. Thus we can
determine a confidence interval from the condition $\Delta\log{\cal
  L}>0.5$, resulting in approximately $165\kmsin \apprla
v_{circ}(7R_e) \apprla 250\kmsin$ at $1\sigma$. This would exclude
both model A without dark matter and the most massive halo model E.
However, we do not believe this is a very strong result, given the
influence of a single outlier on the likelihood values in
Table~\ref{tab:lh}, and the asymmetries in some of the LOSVDs (see
Fig.~\ref{fig:sphlosvd}). Note also that all models are consistent with
the data at the $2\sigma$ level, for which $v_{circ}(7R_e) \apprla
280\kmsin$.

\section{Summary and conclusions}
\label{sec:conclusions}

In this paper, we have carried out a dynamical study of the elliptical
galaxy NGC 3379.  This intermediate luminosity E1 galaxy has a rapidly
declining velocity dispersion profile, which has been taken as
evidence by \citet{romanowsky_etal03} and \citet{douglas+07} that this
galaxy may lack the kind of dark matter halo that the current
$\Lambda$CDM cosmology requires.

To explore this issue further, we have combined photometry, long slit
spectroscopic data, SAURON absorption line kinematics and PN velocity
dispersion data, to fit dynamical models in a sequence of
potentials whose circular velocity curves at large radii vary between
a near-Keplerian decline and the nearly flat shapes generated by
massive halos.  The combined kinematic data set runs from the center
of NGC 3379 to about $7$ effective radii.

For constructing the dynamical models we have used the flexible
$\chi^2$-made-to-measure particle code NMAGIC developed by
\citep{delo+07, delo+08}.  The NMAGIC models described in this paper
consist of $7.5\times10^5$ particles, and for the first time  are
constructed for such a comprehensive data set, including integral
field kinematic data.

We find that a variety of dynamical models both with and without dark
matter produce viable fits to all the data. For assumed spherical
symmetry we find that the data are consistent both with near-isotropic
systems which are dominated by the stellar mass out to the last
kinematic data points, and with models in massive halos
whose outer parts are strongly radially anisotropic ($\beta(7R_e)\apprga
0.8$). In these latter models, the stellar mass distribution dominates
in the center, and the dark matter fraction is $\sim 60\%$ of the
total at $7 R_e$.

In the spherical potentials we have also used the likelihood scheme
of \citet{delo+08} to fit the models directly to the PN velocities.
From the likelihood values obtained in these fits as well as the
posterior likelihoods of the models fit to the dispersion profiles,
we estimate confidence limits on the halo circular velocity at
$7R_e$, resulting in approximately $165\kmsin \apprla
v_{circ}(7R_e) \apprla 250\kmsin$ at $1\sigma$. This would exclude
both the model without dark matter and the most massive halo model
E in our sequence which has $v_{circ}(7R_e) \simeq 275\kmsin$.

To illustrate the power of NMAGIC we have used it to find the shape of
a model flattened along the line-of-sight in a spherical potential
including this most massive halo E, which fits all the kinematic data
with high accuracy. However, all attempts to find more realistic
models with this massive halo have failed, suggesting that we may have
found the upper limit of the range of consistent mass distributions.

We have then constructed self-consistent axisymmetric models of
inclinations $i=90\degrees$, $i=50\degrees$, and $i=40\degrees$ in the
same sequence of halos potentials. These models essentially confirm
the spherical results. The edge-on models are very similar to the
spherical models, becoming highly anisotropic in the more massive
halos. The inclined models in addition become more flattened at large
radii, which helps in decreasing the outer velocity dispersion
profile.  Finally, we have constructed a weakly triaxial model, in the
most massive allowed halo D, by free evolution from an axisymmetric
model flattened along the line-of-sight. This model ended at
$i\simeq45\deg$ inclination, is almost axisymmetric, and matches all
kinematic data very well, similarly to the inclined axisymmetric
models in this halo potential. All these models are stable over Gyrs.

Our main conclusions are as follows:

(i) The kinematic data for NGC 3379 out to $7R_e$ are consistent with
a variety of potentials and do not give strong constraints on the mass
distribution in this galaxy.  The main reason for this is the
well-known degeneracy between mass and radial anisotropy which is
substantial when the velocity dispersion profile falls with
radius. Formal confidence limits on the halo circular velocity at
$7R_e$ are $165\kmsin \apprla v_{circ}(7R_e) \apprla 250\kmsin$ at
$1\sigma$, which would weakly exclude models without dark matter.

(ii) NGC 3379 may well have the kind of dark matter halo consistent
with the current $\Lambda$CDM paradigm. The circular velocity curves
of the merger models constructed by \citet{dekel+05} in the
$\Lambda$CDM cosmology framework,
$v_{circ}(r)/v_{circ}(R_e)\simeq(R/R_e)^{-0.135}$, fall right into the
range of circular velocities of our best-fitting models in
Fig.~\ref{fig:circvel}. 

(iii) Such models, however, are required by the data to have strongly
radially anisotropic orbit distributions in their outer regions,
$\beta\apprga0.8$ at $7R_e$, while model predictions at $7R_e$ are
$0.3\apprla\beta\apprla0.6$ \citep{dekel+05} and $\beta\simeq0.7$
\citep{abadi+06}.  Kinematic data at even larger radii than presently
available would be required to discriminate between these models and
less anisotropic models with lower mass halos.

\bigskip
\noindent

\section*{Acknowledgments}

We thank John Magorrian making his software available, Jens Thomas
for helpful advice and discussions, and the referee for a careful
reading of the manuscript.

\bibliography{mybib} 

\label{lastpage} 

\end{document}